\documentclass[aip,twocolumn,letterpaper]{revtex4}

\usepackage{fullpage}

\usepackage{graphics}
\usepackage{epsfig}

\usepackage{xcolor}

\usepackage[margin=1.0in]{geometry}

\begin{document}
\title{Required toroidal confinement for fusion and omnigeneity}
\author{Allen H Boozer}
\affiliation{Columbia University, New York, NY  10027\\ ahb17@columbia.edu}

\begin{abstract}

Deuterium-tritium (DT) burning requires a long energy confinement times compared to collision times, so the particle distribution functions must approximate local-Maxwellians.   Non-equilibrium thermodynamics is applicable, which gives relations among transport, entropy production, the collision frequency, and the deviation from a Maxwellian.  The distribution functions are given by the Fokker-Planck equation, which is an advection-diffusion equation.  A large hyperbolic operator, the Vlasov operator with the particle trajectories as its characteristics, equals a small diffusive operator, the collision operator.  The collisionless particle trajectories would be chaotic in stellarators without careful optimization.  This would lead to rapid entropy production and transport---far beyond what is consistent with a self-sustaining DT burn.  Omnigeneity is the weakest general condition that is consistent with a sufficiently small entropy production associated with the thermal particle trajectories.  Omnigeneity requires that the contours of constant magnetic field strength be unbounded in at least one of the two angular coordinates in magnetic surfaces and that there be a symmetry in the field-strength wells along the field lines.   Even in omnigenous plasmas, fluctuations due to microturbulence can produce chaotic particle trajectories and the gyro-Bohm transport seen in many stellarator and tokamak experiments.  The higher the plasma temperature above 10~keV, the smaller the transport must be compared to gyro-Bohm for a self-sustaining DT burn.  The hot alphas of DT fusion heat the electrons.  When the ion-electron equilibration time is long compared to the ion energy confinement time, a self-sustaining DT burn is not possible, which sets a limit on the electron temperature.

\end{abstract}

\date{\today} 
\maketitle

\section{Introduction}

Fusion power plants that utilize magnetic confinement will be in a paradoxical plasma collisionality regime \cite{Boozer:RMP}, which produces a number of subtleties.  The smallness of the cross section for the fusion of deuterium (D) and tritium (T) compared to the Coulomb cross section requires the plasma confinement time be long compared to ion and electron collision times and sets the temperature $T$ at which power plants must operate.  Practical limits on the power density from DT fusion, which scales approximately as $(nT)^2$, requires the number density $n$ be sufficiently low that the mean-free-path of electrons and ions between collisions be approximately a thousand times longer than the size of the plasma. 

The shortness of collision compared to confinement times implies the distribution functions of the ions and electrons can have only small departures from Maxwellians.  This makes non-equilibrium thermodynamics applicable, which gives relations among transport, entropy production, the collision frequencies, and the deviations from Maxwellians.  The fundamental description of the particle distribution functions is the Fokker-Planck equation, which is of the advection-diffusion form and paradoxically allows transport coefficients that are independent of the collision frequency.  An example, is gyro-Bohm diffusion, which approximates the empirically observed confinement of tokamaks and stellarators.  Gyro-Bohm transport implies adequate confinement for a self-sustaining fusion burn is more difficult the higher the plasma temperature above 10~keV, Figure \ref{fig:gyro-Bohm}. The plasma heating in DT fusion is through alpha particles heating the electrons, which constrains the electron temperature for which a self-sustaining DT burn is possible, Figure \ref{fig:T_i}.  These topics are discussed in Section \ref{sec:genl-tranp}.

In addition, collisionless particle drifts can be inconsistent with the confinement required for fusion---omnigeneity is the weakest general condition for adequate confinement of thermal particles.  Omnigeneity requires that the contours of constant magnetic field strength be unbounded in at least one of the two angular coordinates in magnetic surfaces and that there be a symmetry in the field-strength wells along the field lines.

Without collisions, passing particles, which have a velocity parallel to the magnetic field, $v_{||}$, that is never zero, require the magnetic field lines lie on toroidal surfaces for confinement.  A particular magnetic surface is denoted by the magnetic flux $\psi$, called the toroidal flux, it encloses.  The magnetic field can be written in the Clebsch form,
\begin{equation}
2\pi\vec{B}=\vec{\nabla}\psi\times\vec{\nabla}\theta_0, \label{Clebsch}
\end{equation}
where a poloidal angle $\theta$ can be defined so $\theta=\theta_0 + \iota(\psi)\varphi$ with $\iota$ the rotational transform and $\varphi$ a toroidal angle, Appendix \ref{Boozer-coord}.

Collisionless trapped particles, which have turning points at which $v_{||}=0$, can be difficult to confine in a toroidal plasma.  Omneigeneity is the weakest condition that ensures that successive turning points of collisionless trapped particles are on the same magnetic surface and that particle trajectories have a maximum deviation from magnetic surfaces that is proportional to their gyroradius.  When the maximum deviation of collisionless particles from magnetic surfaces scales as their gyroradius, their collisional confinement is adequate for a fusion power plant.

The definition of omnigeneity is based on the longitudinal action $J$ of Northrop and Teller \cite{Northrop-Teller:1960};
\begin{eqnarray}
\frac{\partial J}{\partial\theta_0}=0, \hspace{0.1in}\mbox{where}\hspace{0.1in}\\
J(\psi,\theta_0,u)\equiv m\int v_{||}d\ell. \label{J}
\end{eqnarray}
The integral is over the differential distance $d\ell$ along a field line from one turning point to the next of a trapped particle of energy $u$ and mass $m$.  A particular magnetic field line is denoted by a constant $\psi$ and $\theta_0$.  The departure from omnigeneity can be measured by the dimensionless quantity $(\partial J/\partial\theta_0)/J$ as discussed in Section \ref{sec:departure meas}.

The conservation of the action $J$ along the trajectory of a particle, which we call an isoaction trajectory, generally provides adequate confinement for the alpha particles produced by DT fusion but not for the thermal particles.  The difference is that alpha particles slow down on the electrons preserving their initial pitch angle while thermal particles scatter rapidly in pitch compared to their confinement time.

Two types of departures of thermal particles from omnigeneity are consistent with fusion power plants. (1)  As noted in \cite{Boozer:stell-imp}, there are advantages in having rapid transport in the core of a fusion plasma with the required confinement provided by an annulus.  Advantages include (a) obtaining fusion over a larger fraction of the plasma volume, (b) avoiding impurity accumulation in the center, (c) simpler fueling by pellet injection when the annulus is thin enough to be penetrated by pellets, and (d) a method of controlling the microturbulence through control of the density gradient across the annulus. (2) As discussed in Section \ref{action failure}, when the magnetic field strength departs from an omnigenous form by forming secondary magnetic wells only in the immediate vicinity of extrema of the magnetic field strength, the confinement can remain adequate for fusion.  This requires magnetic field strength variations that are both sufficiently weak and short wavelength along $\vec{B}$.

The effects of particle drifts and omnigeneity are  discussed in Sections \ref{sec:drift traj} through \ref{Particle drift}.

Section \ref{sec:drift traj}, \emph{Particle drift trajectories}, discusses the basic features of particle motion in the small-gyroradius limit, which is the relevant limit of fusion power plants.  Results from the study of  weak toroidal ripple in tokamaks are shown to imply that breaking action conservation near maxima and minima of the field strength can have remarkably little effect on transport.

Section \ref{sec:J},  \emph{Derivatives of the action}, expresses the bounce time, precession, and radial drift of trapped particles in terms of derivatives of the action $J$.  These expressions are used to obtain formulas that are integrals of the field strength along the magnetic field lines. 

Section \ref{costruc-omnig}, \emph{Construction of omnigenous equilibria}, discusses the freedom to produce exact omnigeneity on a surface based on the formulas derived in  Section \ref{sec:J}.

Section \ref{Particle drift}, \emph{Effects of trapped particle drifts}, obtains the deviation of a particle from the flux surface on which its turning points are located in an omnigenous system. The bootstrap current is shown to be unchanged by the deviation of an omnigenous stellarator from quasi-symmetry.

Section \ref{Discussion}, \emph{Discussion}, reviews the results of the paper.

Appendix \ref{Boozer-coord}, \emph{Magnetic fields in toroidal plasmas}, is a short review of Boozer coordinates for toroidally confined plasmas.


\section{General transport properties \label{sec:genl-tranp} }


\subsection{Near-Maxwellian requirement \label{Maxwellian-req}}

A power plant based on the fusion of deuterium and tritium requires the confinement of a plasma for hundreds of collision times for the ions and approximately ten thousand collision times for the electrons.  The implication is that the distribution functions $f$ must be very close to local Maxwellians $f_M$.  


\subsubsection{Non-equilibrium thermodynamics \label{sec:non-eq}}

Since the distribution functions of fusion plasmas must be close to Maxwellian, the theory of non-equilibrium thermodynamics has important implications on the solutions to the Fokker-Planck equation \cite{Boozer:RMP,Boozer:NF-rev2015}. 

Writing $f=f_me^{\hat{f}}$,  the ratio of the energy confinement time to the collision time for a species is comparable to $1/\left<\hat{f}^2\right>$ with $<\cdots>$ a velocity space average  \cite{Boozer:RMP}.  More precisely plasma particle or energy transport implies entropy production per unit volume by collisions, 
\begin{equation}
\dot{s}_c =\frac{T}{2m}\int \frac{\partial \hat{f}}{\partial \vec{v}}\cdot\tensor{\nu}\cdot\frac{\partial \hat{f}}{\partial \vec{v}} e^{\hat{f}} f_m d^3v, \label{coll ds/dt}
\end{equation}
where $\tensor{\nu}$ is the tensor collision frequency.  See Equation (146) in \cite{Boozer:RMP}.  For a steady-state burning plasma, the  power from DT fusion must balance the power loss associated with collisional entropy production, 
\begin{equation}
\dot{s}_c=-\Gamma\frac{d\mu_c/T}{d\psi_t} +\mathcal{F}_h\frac{d1/T}{d\psi_t} \label{trans ds/dt}
\end{equation}
with $\mu_c=T\ln(n/T^{3/2})$ the chemical potential, $\Gamma\equiv\vec{\Gamma}\cdot\vec{\nabla}\psi_t$ the particle flux and $\mathcal{F}_h\equiv\vec{\mathcal{F}}_h\cdot\vec{\nabla}\psi_t$ the heat flux.  The quantities $-d(\mu_c/T)/d\psi_t$ and $d(1/T)/d\psi_t$ are called thermodynamic forces; $\Gamma$ and $\mathcal{F}_h$ are their conjugate fluxes.  See Equation (151) of \cite{Boozer:RMP}.   A more detailed discussion of  the application of non-equilibrium thermodynamics and kinetic theory to plasma transport is given in \cite{Boozer:NF-rev2015}.  

Theory of non-equilibrium thermodynamics and kinetic theory \cite{Boozer:NF-rev2015} allow one to address the subtle question of when like-particle collisions cause particle transport. This question is important for determining the ambipolar electric field, which has a large effect on impurity transport.   The basic answer is that like-particle collisions can produce particle transport only through the viscosity tensor, which in fusion plasmas is well approximated by the parallel viscosity.  The details require too much space for this paper.


\subsubsection{The Fokker-Planck equation as an advection-diffusion equation}

Mathematically, the Fokker-Planck equation for the distribution function $f$ is of the advection-diffusion type.  The advective part is called the Vlasov equation, which is hyperbolic in type with characteristics given by the trajectories of the particle Hamiltonians, $H(\vec{p},\vec{x},t)$.  The entropy density $s=-\int f\ln fd^3v$ is an invariant of the Vlasov equation. The diffusive part of the Fokker-Planck equation is the collision operator, which in plasmas has two special properties.  The collision operator (1) diffuses particles in velocity $\vec{v}$ but not spatially, in $\vec{x}$, which makes the entropy production per unit volume well defined, and (2) has no effect on local Maxwellians when $T_e=T_i$ due to the conservation of energy and momentum by collisions.  The ratio of the large Vlasov operator to the small collision operator is $v/L\nu$, which is the ratio of the mean free path to a characteristic spatial scale; $\sim10^3$ in a power plant.

The advective term in advective-diffusion equations generally has chaotic characteristics, which means infinitesimally separated characteristics separate exponentially even while remaining in a bounded volume of phase space.  When the characteristics are chaotic, the effects of diffusion are exponentially amplified.  This is true for mixing in fluids \cite{Aref:1984}, such as temperature equilibration in a room, and in magnetic reconnection \cite{Boozer:reconnecton2023} as well as in the entropy production by solutions to the Fokker-Planck equation, Equation (\ref{coll ds/dt}).   Chaos causes an exponential increase in $\partial\hat{f}/\partial\vec{v}$ until collisions become important.

Two features make solutions of the Fokker-Planck equation that are of relevance to fusion plasma atypical for advection-diffusion equations.  (1) Collisional diffusion vanishes when the distribution is a local Maxwellian.  (2) The required long confinement times compared to collision times of the ions and electrons in fusion plasmas imply the departures, $\hat{f}$, from local Maxwellian distributions must be small.  The diffusion operators for fluid mixing and magnetic reconnection, have only a trivial null space---a constant distribution---unlike the local-Maxwellian null space of the collision operator.

In stellarator fusion plasmas, the external magnetic field must be chosen to have the special property that  the particle Hamiltonians $H(\vec{p},\vec{x})$ have constants of the motion such that when the distribution functions for a confined plasma are written in terms of these constants of the motion the deviations $\hat{f}$ from local Maxwellians are small.  The existence of these constants of the motion precludes the chaotic trajectories which exponentially enhance entropy production. 

The absence of continuous symmetries was thought to preclude appropriate constants of motion and to be a fatal flaw for the use of stellarators for fusion power plants.  It was, therefore, surprising when it was found that such constants of the motion theoretically exist for stellarators \cite{Boozer:quasisymmetry} and even more remarkable that stellarators could be designed \cite{Nuhrenberg-Zille:1988} and operated that exploit this \cite{Beidler:2021}.  

The disproof of the fatal flaw of stellarators is of fundamental importance to the achievement of fusion energy \cite{Boozer:stell-imp}.  Unlike in tokamaks, the properties of the confining magnetic field in stellarators can be dominated by the externally produced magnetic field.   This produces cage-like robust confinement, unique reliability of computational optimization, and steady-state burning plasmas with zero external power input to the plasma.  The absence of continuous spatial symmetries, which made the existence of constants of the motion subtle, implies that feasible coils can produce approximately ten times as many distributions of magnetic field as in an axisymmetric tokamak.  The optimization space is approximately ten times larger in a stellarator than in a tokamak.

When the $\partial\hat{f}/\partial\vec{v}$ given by the constants of the motion can be made small, the neoclassical entropy production is given by Equation (\ref{coll ds/dt}) and the transport can be obtained using Equation (\ref{trans ds/dt}).  

Even when the Vlasov equation deviation $\partial\hat{f}/\partial\vec{v}$ cannot be made small everywhere, the full Fokker-Planck equation can give an $\partial\hat{f}/\partial\vec{v}$, which when substituted into  Equation (\ref{coll ds/dt}) gives sufficiently small neoclassical transport for fusion power.  This is particularly true when the Vlasov $\partial\hat{f}/\partial\vec{v}$ is large in only a small region of velocity space---such as near the limit of  either deeply and barely trapped particles, which is discussed in Section \ref{action failure}.

No matter how well the external magnetic field is designed so $H(\vec{p},\vec{x})$ has appropriate constants of the motion, the plasma itself can produce electric fluctuations that cause the particle trajectories to be chaotic.   When this occurs, the Fokker-Planck equation becomes exponentially sensitive to the diffusivity of the collision operator, as is typical of solutions to the advection-diffusion equation.  This sensitivity allows collisional entropy production, Equation (\ref{coll ds/dt}), even in the limit as collision frequency $\tensor{\nu}$ goes to zero.  GyroBohm diffusion, $D_{gB}\equiv (\rho_i/a)T/eB$, which approximates the commonly observed transport in toroidal plasmas \cite{Boozer:stell-imp}, is an example.  The spatial scale of gradients is $a$ and $\rho_i$ is the ion gyroradius.  

Three types of electric field fluctuations can make the particle trajectories chaotic: fluctuations (1) in the electric potential, (2) in the locations of the magnetic field lines, and (3) in the loop voltage, which breaks the magnetic surfaces.  These can be identified from the general representation of an electric field in terms of a magnetic field \cite{Boozer coord},
\begin{eqnarray}
&& \vec{E}+\vec{u}_\bot\times\vec{B}=-\vec{\nabla}\Phi + \frac{V_\ell}{2\pi}\vec{\nabla}\varphi;\\
&& V_\ell \equiv \lim_{\ell\rightarrow\infty} \frac{2\pi\int\vec{E}\cdot \hat{b}d\ell}{\int \frac{\partial \varphi}{\partial \ell} d \ell}.
\end{eqnarray}
The loop voltage $V_{\ell}$ is a spatial constant along a magnetic field line and measures the breaking of the lines, $d\ell$ is the differential distance a magnetic field line, $\hat{b}(\vec{x},t)=\vec{B}/B$, and $\vec{u}_\bot(\vec{x},t)$ is the flow velocity of magnetic field lines.   Relatively little work has been done on field-line breaking microturbulence.


\subsection{Gyro-Bohm transport and fusion}

\begin{figure}
\centerline{ \includegraphics[width=3.0in]{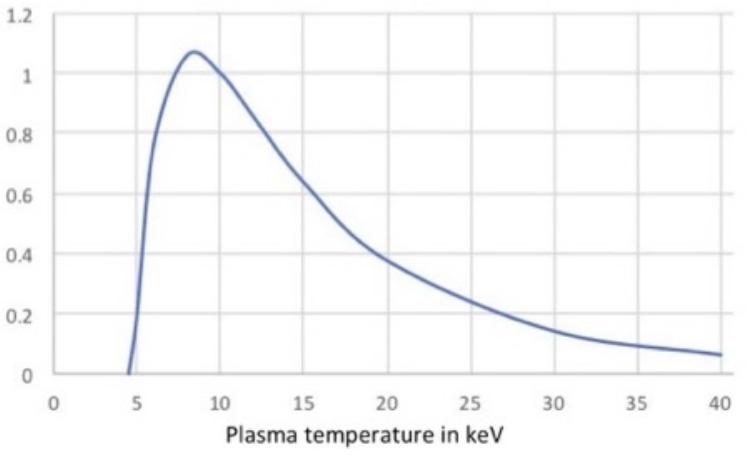} }
\caption{The allowed transport relative to gyro-Bohm for a self-sustaining fusion burn is shown as a function of the plasma temperature.  Gyro-Bohm diffusion gives a confinement-time scaling comparable to that seen in tokamaks and stellarators.  Achievement of a DT burn at a high plasma temperature is difficult.}
\label{fig:gyro-Bohm}
\end{figure}

Since gyro-Bohm transport commonly approximates the confinement time of tokamak and stellarator plasmas \cite{Boozer:stell-imp}, it is important to study the effect on the sustainment of a DT fusion burn.  Figure \ref{fig:gyro-Bohm} is constructed using the values of gyro-Bohm diffusion and of alpha-power production by DT fusion corrected for bremsstrahlung losses that are given in the Appendix to \cite{Boozer:stell-imp} and discussed in Appendix 3 of \cite{Boozer:fast-path}.  The energy confinement need only be comparable to gyro-Bohm at 10~keV, but the confinement must be an order of magnitude better above 35~keV. Stellarator power-plant designs typically have a 10~keV temperature, but issues such as current maintenance and the Greenwald limit on the density can push tokamak power plants to higher temperatures where far better confinement is required \cite{Boozer:fast-path}.

\subsection{Ion and electron energy transport \label{sec:ion-electron consraint}}

\begin{figure}
\centerline{ \includegraphics[width=3.0in]{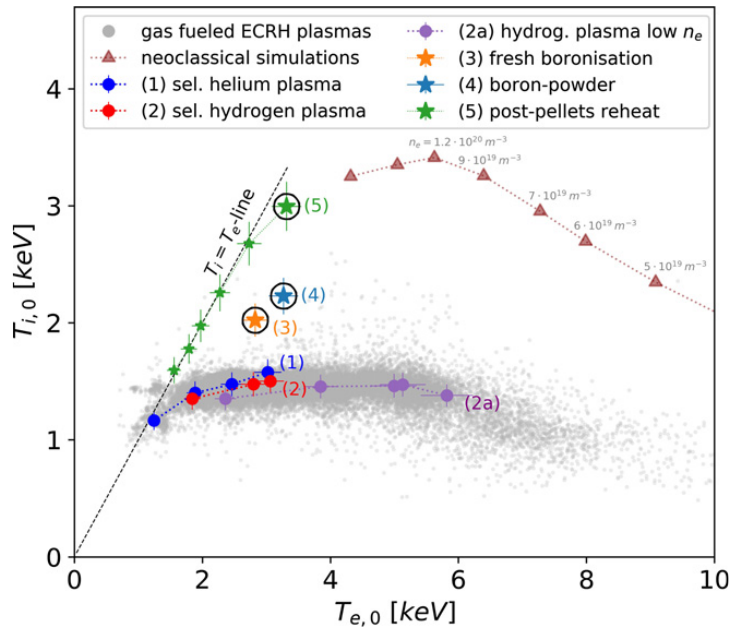} }
\caption{In W-7X experiments, the central ion temperature $T_{i,0}$ is seen to pass through a maximum and then drop as the central electron temperature $T_{e,0}$ is increased.  Computer simulations that assume neoclassical transport show a similar effect but with a higher maximum for the ion temperature.  This figure is copied with permission from M. N. A Beurskens, S. A. Bozhenkov, O. Ford, et al.,  Nucl. Fusion \textbf{61} 116072 (2021).  Achievement of a self-sustaining DT burn is difficult when $T_e$ is sufficiently high that the electron-ion equilibration-time is long compared to the ion energy-confinement time.  }
\label{fig:T_i}
\end{figure}

In self-sustaining fusion and in electron-cyclotron heated plasmas, the heating power goes into the electrons and only by equilibration heats the ions.   When the external heating heats only the electrons, and the electron-ion equilibration-time, $\tau_{eq}\propto T_e^{3/2}$, is long compared to the ion energy confinement  time, $\tau_{Ei}$, the ions are colder than the electrons, by the ratio $T_i/T_e=3\tau_{Ei}/2\tau_{eq}$.  The electron temperature can then rise to whatever level is required for their energy confinement time $\tau_{Ee}$ to balance the input power.  For a self-sustaining DT fusion burn, neither the ion energy confinement time nor the electron energy confinement can be significantly shorter than the plasma confinement time required for a DT burn $\tau_{DT}$.   More remarkably, the electron energy confinement time cannot be substantially longer. 

When the plasma heating is electron heating, the steady-state ion temperature $T_i$ is 
\begin{eqnarray}
\frac{\frac{3}{2}T_i}{\tau_{Ei}} &=& \frac{T_e-T_i}{\tau_{eq}} \hspace{0.2in}\mbox{or   }\\
\frac{T_i}{T_e} &=&\frac{\tau_{Ei}}{\tau_{Ei}+\frac{2}{3}\tau_{eq}}
\end{eqnarray}

The steady-state electron temperature $T_e$ depends on the heating power per particle $p_h$:
\begin{eqnarray}
p_h &=& \left(\frac{1}{\tau_{Ee}} +\frac{1}{\frac{2}{3}\tau_{eq}}\right)T_e -\frac{T_i}{\frac{2}{3}\tau_{eq}} \hspace{0.2in}\mbox{so  }\\
&=&  \left(\frac{1}{\tau_{Ee}} + \frac{1}{ \tau_{Ei}+\frac{2}{3}\tau_{eq}} \right)T_e. \label{T_e eq}
\end{eqnarray}
When the equilibration time is long compared to the ion energy confinement time, the power going into the ions drops as $1/\sqrt{T_e}$, and at fixed input power, the only limit on the electron temperature is the electron energy confinement time.  When $p_h$ is a DT fusion burn, the power input per particle is approximately $p_{DT}=T_i^2/(T_{DT}\tau_{DT})$, where $T_{DT}\approx 10~$keV is the optimal temperature for a burn, Figure \ref{fig:gyro-Bohm}, and $\tau_{DT}$ is the required time for a DT burn at $T_{DT}$.

The equilibration time is given by $n\tau_{eq} \approx 0.5 T_e^{3/2}$, and the requirement for DT ignition was given in Equation (A6) of Reference \cite{Boozer:stell-imp} as $nT_{DT}\tau_{DT}\approx 3$.  Densities are measured in units of $10^{20}/~$m$^3$ and temperatures in  units of $10$~keV.   The equilibration time becomes longer than $\tau_{DT}$ for $T_e\gtrsim30$~keV.


\subsection{Importance of omnigeneity}

Although magnetically-confined fusion plasmas must be sufficiently collisional for the distribution functions to be near-Maxwellian, in another sense magnetically confined fusion plasmas must be almost collisionless.  The mean free path between collisions for ions and electrons are both of order 10~km.  As discussed in Section \ref{Maxwellian-req}, it is the sensitivity of advection-diffusion equations to chaos when diffusion is small that makes the existence of constants of the motion so important.  Omnigeneity, $\partial J/\partial\theta_0=0$, is the weakest general condition for obtaining an appropriate constant of the motion.  The conservation of the action $J$ alone is generally inadequate.

Ideally, the distribution functions would be Maxwellians with the number density $n$ and temperature $T$ functions of the toroidal magnetic flux $\psi$.  But, collisonless particle trajectories make excursions $\Delta\psi$ away from their home $\psi$-surfaces.  The $\Delta\psi$ of collisionless trajectories can either be proportional to the gyroradius, $\rho\equiv mv/qB$, of the particle or independent of the gyroradius as $\rho\rightarrow0$.

When the excursions $\Delta\psi$ of all particles in a plasma of radius $a$ are proportional to the gyroradius the deviations from local Maxwellians scale as $\hat{f}\propto \rho/a$, which is adequate for fusion power plants.

When the excursions $\Delta\psi$  of a fraction of the particles is independent of their gyroradius, the effect on transport---particularly for high-energy $\alpha$ particles and for electrons at low collisionality---can be inconsistent with fusion power plants. In the low collisionality limit, the intrinsic electron transport becomes larger than the intrinsic ion transport by the ratio of their collision frequencies when $\Delta\psi$ is independent of the gyroradius. 

When the intrinsic electron particle transport is more rapid than the ion, an electric potential $\phi(\psi)$ is required to confine the electrons.  This has the advantage of expelling impurities.  However, the maximum collisionality for this limit is lower than the collisionality in most designs for fusion power plants.  In  stellarator power plant designs, the electrons are generally better confined than the ions, which implies a $\phi(\psi)$ that confines ions and impurities.

Careful design of stellarators is required to obtain a scaling of the excursions $\Delta\psi$ that is proportional to the gyroradius.  The least constraining design principle that is consistent with this scaling is omnigeneity, which makes this concept of central importance to magnetic confinement fusion.


\section{Particle drift trajectories \label{sec:drift traj} }

\subsection{General theory for collisionless particles}

In 1980, Boozer \cite{Boozer:1980} showed that as  $\rho\rightarrow0$, the center of the circle about which a particle gyrates has a drift velocity 
\begin{eqnarray}
\vec{v}_d &=& \frac{v_{||}}{B}\vec{H}; \label{v_d} \label{v_d eq}\\
\vec{H}&\equiv& \vec{B} +\vec{\nabla}\times( \rho_{||} \vec{B}).
\end{eqnarray}
The \emph{parallel gyroradius} $\rho_{||} \equiv mv_{||}/B$ is a function of the position $\vec{x}$ of the particle and its two conserved quantities
\begin{eqnarray}
&& \mu\equiv\frac{mv_\bot^2}{2B} \hspace{0.3in} \mbox{the magnetic moment, and}\hspace{0.2in}\\
&&u\equiv\frac{1}{2}mv_{||}^2 + \mu B +q\phi \hspace{0.3in} \mbox{the energy.} \label{energy-eq}
\end{eqnarray}
The magnitude of the particle velocity parallel and perpendicular to $\vec{B}$ are $v_{||}$ and $v_\bot$.  When the energy confinement time of a toroidal plasma is long compared to the collision times, the electric potential $\phi(\vec{x})$ has only small departures from being constant along magnetic field lines.

Energy conservation for particles,  $du/dt=0$, holds when the system is time independent.  Magnetic moment conservation, $d\mu/dt=0$, requires the magnetic field experienced by the particle change little from one gyro-period to the next.  This requires $\rho_{||}/L<<1$, where $L$ is the scale of variations in the  magnetic field in the direction along $\vec{B}$.  

For passing particles, particles that have a parallel velocity that is never zero, $v_{||}\neq0$, confinement is generally assured by the existence of magnetic surfaces that are denoted by $\psi$, the toroidal magnetic flux enclosed by the surface.   Wherever $v_{||}\neq0$ the particles follow the effective magnetic field $\vec{H}$, which differs from $\vec{B}$ by only a small perturbation.  Consequently, passing particles, those that never have $v_{||}=0$, are generally well confined.  

$\vec{H}$ is singular wherever $v_{||}=0$, which are the turning points of trapped particles.  Trapped particles bounce back and forth in the energy well produced by a variation in the field strength along the magnetic field strength, Equation (\ref{energy-eq}). The drift velocity $\vec{v}_d$ is not singular at $v_{||}=0$ but need not vanish, but does vanish for the special case that $v_{||}=0$ at a maximum or minimum of the field strength.   The singularity of the effective magnetic field $\vec{H}$ can cause trapped particles to drift across the magnetic field lines an arbitrarily large distance.   For a trapped particle, the confinement of the magnetic field lines to constant-$\psi$ surfaces is only relevant to its trajectory between turning points.  The turning points need not remain on a fixed $\psi$-surface as is required for adequate confinement for fusion.  

Trapped-particle confinement can be ensured by either a conserved canonical momentum, which is discussed in Section \ref{sec:quasisymmetry}, or the  longitudinal action $J$, which is discussed in Section \ref{sec:omnngeneity}.  But, $J$-conserving trajectories can have sufficiently large excursions in $\psi$ to strike the walls.  In certain cases in which a short wavelength ripple breaks $J$-conservation, the trajectory excursions directly caused by the breaking can be small, Section \ref{action failure}.

Assuming the electric potential $\phi$ is constant on $\psi$-surfaces, the magnetic field strength at which a particle has a turning point is
\begin{eqnarray}
B_t(\psi) &=& \frac{u-q\phi(\psi)}{\mu}  \label{B_t def}\\
&=&\frac{mv^2}{2\mu}. \label{B_t value}
\end{eqnarray} 
Energy $u$ and magnetic moment $\mu$ conservation imply that $B_t$ is constant of the motion, $dB_t/dt=0$, when the particle remains within a $\psi$-surface.

The most elementary concept for trapped particle confinement is that small gyroradius trapped particles must always lie in a range of $\psi$ such that 
\begin{equation}
B_{max}(\psi)>B_t>B_{min}(\psi). \label{u constraint on psi}
\end{equation}  
When $B_t$ is greater than the maximum field strength on the surface, the particle becomes passing and is thereby well confined. When $B_t$ is less than the minimum field strength on the surface, there is no place the particle can be located on the surface and have positive kinetic energy.

Unfortunately, the constraint of Equation (\ref{u constraint on psi}) on the range of $\psi$ that a charged particle can cover is too broad to be consistent the the confinement of a plasma that has distribution functions that  are sufficiently close Maxwellian for a fusion power plant.

There are four related concepts for obtaining far better trapped particle confinement in toroidal plasma equilibria:  isoaction, omnigeneity, quasi-symmetry, and axisymmetry.  All axisymmetric equilibria are quasi-symmetric, all quasi-symmetric equilibria are omnigenous, and all omnigneous equilibria are isoaction.


\subsection{Symmetry and quasi-symmetry \label{sec:quasisymmetry}}

The most restrictive principle for obtaining trapped particle confinement in a toroidal plasma is axisymmetry, which means $\vec{B}$ is symmetric in the toroidal angle $\varphi$.  The Hamiltonian for charged particle motion has a canonical momentum $p_\varphi =mv_\varphi + q A_\varphi$, which is conserved when the magnetic and electric fields have no $\varphi$ dependence.  This constraint limits excursions of particles from their home surface to distance no greater than the gyroradius in the magnetic field component given by toroidal component of the vector potential, $A_\varphi$, which is the poloidal field.

The toroidal symmetry must be broken for a magnetic field that is curl-free throughout a toroidal region to produce a poloidal magnetic field, which is the definition of a stellarator.  Absence of toroidal symmetry implies the $p_\varphi$ of the exact particle Hamiltonian is not conserved, so it cannot be used to ensure particle confinement in a stellarator.  

When the gyroradius $\rho$ of a particle is small, the center of the circle about which the particle gyrates moves with the drift velocity $\vec{v}_d$, given by Equation  (\ref{v_d eq}) and this velocity is given by a Hamiltonian, called the drift Hamiltonian  \cite{Boozer:drift-H}, which has a simple representation in Boozer coordinates, Appendix \ref{Boozer-coord}.  

The remarkable feature of the drift Hamiltonian is that if toroidal symmetry is broken but the magnetic field strength and the electric potential are functions of $\psi$ and 
\begin{equation}
\zeta=N\varphi - M\theta \label{zeta def}
\end{equation} 
with $N$ and $M$ mutually prime integers, then a canonical momentum of the drift Hamiltonian, $H_d$, is conserved, $dP_h/dt=0$.
\begin{eqnarray}
P_h &\equiv& N P_\theta + MP_\varphi \\
&=&\mu_0\frac{NI+MG}{2\pi B} mv_{||} + q\frac{N\psi - M\psi_p(\psi)}{2\pi}. \hspace{0.3in}\\
\frac{d P_\theta}{dt} &=& - \frac{\partial H_d}{\partial\theta} \hspace{0.2in}   \mbox{and}  \hspace{0.2in} 
\frac{d P_\varphi}{dt} = - \frac{\partial H_d}{\partial\varphi}. 
\end{eqnarray}
The magnetic field is and will be assumed to be given in Boozer coordinates, Appendix \ref{Boozer-coord}, with $d\psi_p/d\psi=\iota$, the rotational transform. 

Stellarators that obey this symmetry $B(\psi,\zeta)$ are called quasi-symmetric.


\subsection{Trapped particles and the action \label{sec:action-def}}

The theory of trapped-particle confinement in toroidal plasmas is essentially the same as the theory of  the radial confinement of particles in a mirror machine.  

Hall and McNamara \cite{Hall-McNamara} introduced the concept of omnigeneity in 1975 as a way to control radial particle motions in mirror machines, where the magnetic field can be written in the Clebsch representation, Equation (\ref{Clebsch}).  Particles that are trapped in a magnetic mirror have three conserved quantities, $\mu$ the magnetic moment, $u$ the energy, and $J(\psi,\theta_0,u)$ the longitudinal action, Equation (\ref{J}), of Northrop and Teller \cite{Northrop-Teller:1960};  $\psi$ is the magnetic flux, and $\theta_0$ is a polar angle.


\subsubsection{Action conservation}

Action conservation, $dJ/dt=0$, requires that the magnetic field experienced by the particle  change little between successive bounces of a trapped particle. As will be shown, the typical time required from one bounce to the next of a trapped particle, $\tau_b$, is $\approx L_p/(\sqrt{\epsilon} v)$, where $L_p$ is the length of a period of the stellarator and $\pm\epsilon$ is the fractional variation in the field strength.  But, the bounce time $\tau_b$ becomes logarithmically infinite as the turning point of a trapped particle approaches a maximum of the magnetic field.  The drift velocity $\vec{v}_d$ in the magnetic field has the typical value $(\rho/R)v$ in both directions across a magnetic field line, where $R$ is the major radius.  The component of $\vec{v}_d$ that is within the $\psi$ surface is called the precession velocity, which gives the frequency $d\theta_0/dt = \omega_{pr}$ with which the Clebsch angle precesses.    When $\rho/L_p<<1$, the action is accurately conserved except when a particle makes a sudden change in its turning points. 

Action conservation is broken when the conserved field strength, $B_t$, at which a particle has a turning point goes from being less than to exceeding a local maximum of the field strength.  The implication is that contours of constant field strength in a magnetic surface, Section \ref{sec:B contours}, must have an infinite extent in one of the the angular coordinates, so a particle can precess forever and never have its turning point disappear.


\subsubsection{Contours of magnetic field strength \label{sec:B contours}}

When the magnetic field strength within a magnetic surface is known in the form $B(\theta,\varphi)$, the contours of constant field strength are defined by $dB/ds=0$ and are given by the differential equations
\begin{eqnarray}
\frac{d\theta}{ds} &=& - \frac{\frac{\partial B}{\partial\varphi}}{\sqrt{ \left(\frac{\partial B}{\partial\theta}\right)^2+\left(\frac{\partial B}{\partial\varphi}\right)^2}}; \\
\frac{d\varphi}{ds} &=& + \frac{\frac{\partial B}{\partial\theta}}{\sqrt{ \left(\frac{\partial B}{\partial\theta}\right)^2+\left(\frac{\partial B}{\partial\varphi}\right)^2}}.
\end{eqnarray}
To ensure the magnetic field strength at all turning points $B_t$ never passes through a local maximum of $B$ as the particle precesses in $\theta_0=\theta -\iota\varphi$, these contours cannot close at finite values of both $\theta$ and $\varphi$, but must extend to infinity in one or both angles.  

In quasi-symmetry, $B(\psi_t,N\varphi-M\theta)$, the angle $\theta = \theta_s+ Ns/\sqrt{M^2+N^2}$ and the angle $\varphi=\varphi_s -Ms/\sqrt{M^2+N^2}$, so $M\theta-N\varphi=M\theta_s-N\varphi_s$ is constant for $0\leq s \leq \infty$.


\subsection{Omnigenous trajectories \label{sec:omnngeneity}}

When $J(\psi,\theta_0,u)$ is conserved, $dJ/dt =( \partial J/\partial\psi)(d\psi/dt)+ ( \partial J/\partial\theta_0)(d\theta_0/dt)=0$.  Consequently, the radial excursion of a particle as it precesses in $\theta_0$ is 
\begin{eqnarray} 
&& \frac{\partial\psi}{\partial\theta_0}\equiv \psi'(\psi,\theta_0,u) \hspace{0.3in} \mbox{   with  }\\
&& \psi'(\psi,\theta_0,u)=- \frac{ \frac{\partial J}{\partial\theta_0} }{ \frac{\partial J}{\partial\psi}} \label{d psi/d theta}.
\end{eqnarray}
When the excursion $\Delta\psi$ from the home flux surface of a particle is small compared to the $\psi$-scale of the variation of $J$, the excursion is the maximum minus the minimum value of the integral $\int\psi' d\theta_0/2$.
The motion of a particle is omnigenous when $\partial J/\partial\theta_0=0$, and only then is the scaling of radial excursions proportional to the gyroradius.  Equation (\ref{psi-variation}) implies the excursion of a particle from a constant-$\psi$ surface is indeed proportional to its gyroradius $\rho$ when the omnigenous constraint is satisfied.  

When $\rho/R<<1$, where $R$ is the major radius of the torus, omnigeneity requires:

\begin{itemize}
\item The contours of constant-$B$ must be unbounded in at least one of the angles.
\end{itemize} 

 \begin{itemize}
\item The net drift in the $\psi$ direction must be zero between successive turning points of trapped particles.
\end{itemize}

The importance of omnigeneity to toroidal plasmas was recognized and used in the design of the W7-X stellarator \cite{Nuehrenberg:2010}.   In 1997, Cary and Shasharina \cite{Cary:1997} pointed out constraints associated with omnigeneity.  In particular, they showed that although an omnigenous field can be smooth it cannot be analytic, for the angular derivatives can not be continuous to all orders at maxima of $B(\theta,\varphi)$ as required for analyticity, Section \ref{sec:B_max omnig}.   Some deviation from exact omnigeneity is acceptable in fusion systems.  Both the design of quasi-omnigenous magnetic configurations and the adequacy of their confinement require extensive computations.  

A close approximation to omnigeneity provides an important goal in codes that optimize non-axisymmetric systems for magnetic fusion, in particular stellarators.


\subsection{Isoaction trajectories \label{isoaction} }

Isoaction means the action is conserved, $dJ/dt=0$, along the trajectory of a particle, but two successive turning points can lie on magnetic surfaces separated by an amount $\delta\psi$, which is proportional to the gyroradius.  The time reversal invariance of a time-independent drift Hamiltonian implies the particle on its return bounce will drift in the same direction, so its turning point over the full back and forth motion is $2\delta\psi$.  

The drift away from flux surface, on which the turning point lies, accumulates until the precession, $\omega_{pr}\equiv d\theta_0/dt$, carries the particle to a location in which $\psi$-drift has the opposite sign.  That the $\psi$-drift changes sign is implied by the Hamiltonian $u(\psi,\theta_0,J)$.   As noted in Section \ref{sec:J} and proven in \cite{Boozer:RMP}, the particle energy $u(\psi,\theta_0,J)$ is a Hamiltonian with a canonical momentum $q\psi$ and a canonical coordinate $\theta_0/2\pi$.   
 
Energy $u=mv^2/2 +q\phi$ and $\mu$ conservation imply the action can be written as 
\begin{eqnarray}
J&=&mv \int\sqrt{1-\frac{B}{B_t}}d\ell,  \label{simp-J}\\
\end{eqnarray}
where turning-point field, $B_t$, is defined in Equation (\ref{B_t def}).  When the effect of the electric potential $\phi(\psi)$ is negligible, as it is for particles of sufficiently high energy, $B_t$, the turning point field strength, is a constant of the motion.  Although the magnitude of action differs, the contours of constant action in $\psi,\theta_0$ space are themselves then independent of the energy, mass, and charge of the particle.   

The electric potential in a plasma is typically $|\phi|\sim T/e$, and the expression for the action is modified so $B_t$ becomes a function of $\psi$ and charge; $B_t =(u-q\phi(\psi))/\mu$, rather than being a charge-independent constant of the motion, as when $\phi=0$.  The constant-$J$ contours in $\psi,\theta_0$ depend on the energy/temperature ratio, $u/T$, and on the sign of the charge but not on the particle mass.  At equal temperatures, $u/T$ is the same for electrons and hydrogenic ions taken from a Maxwellian.  The magnetic field variation in most toroidal plasmas, $2\epsilon = \Delta B/B$, is significantly less than unity, which makes the energy at which the effects of $\phi$ of the constant-action contours negligible when $u>>T/\epsilon$.

At a given energy, $u$, the precession speed $\sim \rho v/R$ and the constant action surfaces are similar for electrons and hydrogenic ions.  Nevertheless, the implications for transport are very different because of the electron collision frequency is approximately two orders of magnitude larger than the ion.  In power-plant-like plasmas, near-thermal electrons are too collisional to follow their constant-action contours to completion while ions can.  When $\omega_{pr}^2<\nu_e\nu_i$, the product of the two collision frequencies, the electrons are the better confined species.

Constant-action trajectories were studied in asymmetric mirrors \cite{Gibson:1963} in 1963 and is closely related to the concept of banana drift transport \cite{Banana drift}.  

Burby, MacKay, and Naik have studied  the general conditions for avoiding changes in trajectory type \cite{isodrastic}, which breaks action conservation, and call such equilibria isodrastic.  As summarized in an April 2023 email from Joshua Burby: ``\emph{Isodrasticity does not prevent drift away from flux surfaces in general. It does not directly control cross surface drift like omnigeneity does. In fact isodrasticity doesn't refer to flux surfaces at all.  Among conditions that prevent separatrix crossing, isodrasticity is weaker than omnigeneity. (Isodrasticity is the weakest such condition.) But among conditions that prevent first-order bounce-averaged radial drift, omnigeneity is the weakest possible condition. Moreover, the condition that all particles stay within a gyroradius of the flux surface they are born on is equivalent to omnigeneity.}"


\subsection{Failure of action conservation \label{action failure} }

The most difficult place to have action conservation is near the maxima and the minima of the magnetic field strength in the magnetic surfaces.  When the magnetic field strength is an analytic function of the spatial coordinates, exact omneigeneity cannot be achieved near field maxima, Cary and Shasharina \cite{Cary:1997} and Section \ref{sec:B_max omnig}.  When short wavelength variations in the magnetic field strength are present, a different issue generally breaks not only omneigeneity but also action conservation near both the maxima and the minima of the field strength.

The archetypal example of the breaking of action conservation is the short wavelength variation in the magnetic field strength due to the $N_c\approx20$ toroidal field coils of a tokamak.  For a review, see \cite{Catto:2019}; more recent articles with extensive references to the literature are \cite{Parra et al:2022,Paul:2022}.  The magnetic field strength in a tokamak can approximated as 
\begin{equation}
B=B_0\Big(1+ \epsilon_t \cos\theta +\epsilon_c \cos(N_c\varphi)\Big),
\end{equation} 
where $\epsilon_t =r/R$ is the inverse aspect ratio, which is orders of magnitude larger than $\epsilon_c$, the field strength variation, or ripple, due to the $N_c$ toroidal field coils.  

All two-helicity expressions for the magnetic field strength have related transport properties \cite{Boozer:quasisymmetry}, so any quasi-symmetry broken by a short wavelength perturbation behaves similarly to a tokamak with toroidal ripple.   

 Although $\epsilon_c<<\epsilon_t$, it produces secondary magnetic wells as long as $\epsilon_c>(\iota/N_c)^2\epsilon_t$; in a typical tokamak $(N_c/\iota)^2\sim1600$.  These secondary wells exist only close to the maximum and minima produced by $\epsilon_t$ when $\epsilon_c<<(\iota/N_c)\epsilon_t$.  Particles make many bounces, $\sim \iota\sqrt{\epsilon_t}(a/\rho)$ even in the shallowest wells, and break action conservation as their precession takes them in and out of the ripple wells that persist near the maxima and minima.  

When the ripple is strong, $\epsilon_c\gtrsim(\iota/N_c)\epsilon_t$, secondary magnetic wells are produced all long the magnetic field lines---even at $\theta=\pm\pi/2$ where the radial magnetic drift is large and the precession small.  The result is an unacceptable loss of ripple trapped particles.  However, when $(\iota/N_c)\epsilon_t >> \epsilon_c\gtrsim(\iota/N_c)^2\epsilon_t$, the secondary wells only exist near the maxima and minima of the $\epsilon_t$ where the radial magnetic drift is small and the precession large.  Even so, particles trapped in these shallow wells break action conservation as their precession takes them in and out of the ripple wells.  

Although the breaking of action conservation increases the distance particles can depart from their home flux surface, the breaking that occurs for $\epsilon_c\approx(\iota/N_c)^2\epsilon_t$ appears to have minimal effects as $N_c\rightarrow\infty$.   The small $\epsilon_c$ regime is not discussed in the review of ripple losses \cite{Catto:2019}, which indicates a perception of its unimportance.  Nevertheless, it is important to study the importance of the weak ripple regime, since action-breaking short wavelength ripple-like magnetic wells are difficult to avoid near maxima and minima of an otherwise quasi-symmetric or omnigenous magnetic field strength.  Indeed, the existence of non-zero $\epsilon_c$-like effects limits the effort that should be placed on approximating omnigeneity near field maxima.

It should be noted that magnetic field ripple effects do not necessarily result from having space between the coils.  For example, a helically symmetric stellarator can be supported by two, or even one, helical wires.  A canonical momentum is then exactly conserved, and coil ripple does not cause enhanced neoclassical transport.   Another example is the poloidal field coil that produces the X-point in a tokamak with a divertor.  This coil does not break axisymmetry, which implies it does not produce effects associated with coil ripple.

 The design of coils that have minimal ripple transport while having the largest possible space between the coils for access is important but relatively unexplored.  In quasi-helical symmetry, the basic concept is that the helical path of the coils should be that of the magnetic field strength, $B(\psi,\theta-N\varphi)$.


\section{Derivatives of the action  \label{sec:J}}


The derivatives of the action that require differentiation of the magnetic field are subtle because consistent coordinates must be used.   Section \ref{J-mag-coord} explains how this is done using Boozer coordinates,  Appendix \ref{Boozer-coord}, which simplify transport calculations. 

One derivative of the action does not involve the magnetic field.  That derivative is with respect to the energy $u$ and  gives the time required for a trapped particle to bounce between its two turning points,
\begin{eqnarray}
\tau_b&\equiv&\int \frac{d\ell}{v_{||}} \\
&=& \frac{\partial J}{\partial u}.  \label{partial J/ partial u}
\end{eqnarray}

The action $J$ is a function of $(\psi,\theta_0,u)$, which implies the energy $u$ can be written as a function $u(\psi,\theta_0,J)$.  As shown in Section VI.D.4 of \cite{Boozer:RMP}, $u(\psi,\theta_0,J)$ is a Hamiltonian for $d\psi/dt$ and $d\theta_0/dt$ with a canonical momentum $q\psi$ and a canonical coordinate $\theta_0/2\pi$.   This canonical form gives a constraint on the action-conserving drifts. 


\subsection{ $J$ in magnetic coordinates \label{J-mag-coord}}

When the magnetic field strength $B(\psi,\theta,\varphi)$ is given in Boozer coordinates, the infinitesimal distance along the magnetic field lines $d\ell$ must be given in those coordinates as well.  The position vector in Boozer coordinates is $\vec{x}(\psi,\theta_0,\varphi)$.  A magnetic field line in the field $2\pi \vec{B}=\vec{\nabla}\psi\times\vec{\nabla}\theta_0$ has a fixed $\psi$ and $\theta_0=\theta-\iota\varphi$.  Equation (\ref{cov-theta_0}) for $\vec{B}$ and the orthogonality relations of general coordinates,  Appendix \ref{Boozer-coord}, imply
\begin{eqnarray}
\frac{d\ell}{d\varphi}&=&\frac{\vec{B}}{B}\cdot \left(\frac{\partial\vec{x}}{\partial\varphi}\right)_{\psi\theta_0} \\
&=&\frac{\mu_0(G+\iota I)}{2\pi B}. \label{d ell / d varphi}
\end{eqnarray}
Using Equation (\ref{simp-J}) the action $J(\psi,\theta_0,u)$ can be written in magnetic coordinates  as
\begin{eqnarray}
J &=& \sqrt{2mu}\frac{\mu_0(G+\iota I)}{2\pi} \int \sqrt{1-\frac{B}{B_t}} \frac{d\varphi}{B}.\label{full J}
\end{eqnarray}


\subsection{Departure from omnigeneity}

The most important derivative of the action for a discussion of omnigeneity is $\partial J/\partial\theta_0$, which must be zero when omnigeneity holds.  

The derivative
\begin{eqnarray}
&&\frac{d\left(\sqrt{1-\frac{B}{B_t}}/B\right)}{dB}=-\frac{1-\frac{B}{2B_t} }{B^2\sqrt{1-\frac{B}{B_t} } },  \mbox{   so  }\\
&&\frac{\partial J}{\partial\theta_0} = -\sqrt{2mu}\frac{\mu_0(G+\iota I)}{2\pi}\int\frac{1-\frac{B}{2B_t} }{B^2\sqrt{1-\frac{B}{B_t} } }\frac{\partial B}{\partial\theta_0} d\varphi. \label{J-theta_0-phi} \nonumber\\
\end{eqnarray}

\subsubsection{Simple measure of departure \label{sec:departure meas} }

A simple measure of the departure from exact omnigeneity on a magnetic surface for use in optimization calculations is the dimensionless quantity
\begin{eqnarray}
\frac{1}{J}\frac{\partial J}{\partial\theta_0}  = - \frac{\int\frac{1-\frac{B}{2B_t} }{B^2\sqrt{1-\frac{B}{B_t} } }\frac{\partial B}{\partial\theta_0} d\varphi }{ \int \sqrt{1-\frac{B}{B_t}} \frac{d\varphi}{B} }.  \label{deviation}
\end{eqnarray}
$B_t$ is a constant, the magnetic field strength at the turning points of the particle, Equation (\ref{B_t value}), $\theta_0$ is a constant that gives the trajectory of the particle along a magnetic field line between bounces, and the $\varphi$ integration is between two successive turning points.  Given the field strength in Boozer coordinates $B(\psi,\theta,\varphi)=B(\psi,\theta_0+\iota\varphi,\varphi)$, exact omnigeneity on the magnetic surface $\psi$ requires the expression for $(\partial J/\partial\theta_0)/J$ of Equation (\ref{deviation}) be zero for all $B_t$ and $\theta_0$ such that $B_t>B$.  The integration can be started from an arbitrary $\varphi$ location by choosing the initial $\theta$ such that $B_t>B$ and integrating forward and backward to the two $\varphi$ locations where $B=B_t$.

The constant-$J$ radial excursions are given by $(\partial J/\partial\theta_0)/(\partial J/\partial\psi)$ and this ratio has been used in optimization studies \cite{Bader:2019} by minimizing 
\begin{equation}
\gamma_c\equiv \frac{2}{\pi}\arctan\left(\frac{\partial J/\partial\theta_0}{\partial J/\partial\psi}\right).
\end{equation}
When the electric potential $\phi$ satisfies $d\phi/d\psi=0$, $\partial J/\partial\psi$ generally passes through zero between deeply trapped and barely trapped particles.  When $d\phi/d\psi\neq0$, the derivative $\partial J/\partial\psi$ is always zero for a certain kinetic energy for either electrons and ions.   
 
Dividing $\partial J/\partial\theta_0$ by $J$ corrects for the intrinsic smallness of $J$ and its derivatives for deeply trapped particles without introducing the subtleties associated with the vanishing of $\partial J/\partial\psi$ in particular parts of velocity space.

\subsubsection{$\mathcal{S}(\psi,\theta_0,B_t)$ as a measure}

Another measure of the departure from omnigeneity $\mathcal{S}(\psi,\theta_0,B_t)$ will be used in Sections  \ref{sec:B_min omnig} and \ref{sec:B_max omnig} to construct omnigenous fields. 

The integration variable in Equation (\ref{J-theta_0-phi}) can be changed from $d\varphi$ to $dB/(\partial B/\partial\varphi)$.  Then,
\begin{eqnarray}
\frac{\partial J}{\partial\theta_0} &=& \frac{\mu_0(G+\iota I)}{2\pi N_p}\frac{\sqrt{2mu}}{B_t} \mathcal{S}(\theta_0,B_t),  \mbox{   where   }  \label{J-theta_0-S}\\
\mathcal{S}&\equiv&- N_p B_t \int_{B_{min}}^{B_t}\frac{1-\frac{B}{2B_t} }{B^2\sqrt{1-\frac{B}{B_t} } }\frac{ \frac{\partial B}{\partial\theta_0} }{ \frac{\partial B}{\partial\varphi} } dB, \label{J-theta_0-B}
\end{eqnarray}
where $N_p$ is the number of periods of the stellarator.

The integrand in Equation (\ref{J-theta_0-B}) can be rewritten using
\begin{eqnarray}
&&\frac{ \left(\frac{\partial B}{\partial \theta_0}\right)_\varphi }{\left(\frac{\partial B}{\partial \varphi}\right)_{\theta_0}}=-\left(\frac{\partial\varphi}{\partial\theta_0}\right)_B, \mbox{   which follows from  } \hspace{0.2in}\\
&& dB = \frac{\partial B}{\partial\psi} d\psi +\frac{\partial B}{\partial\theta_0} d\theta_0 + \frac{\partial B}{\partial\varphi} d\varphi.
\end{eqnarray}

There is a subtlety in Equation (\ref{J-theta_0-B}) that must be addressed. At each value of $B$ in the range $B_t\geq B \geq B_{min}$ represents at least two locations along a magnetic field line.  Regions in which $(\partial B/\partial\varphi)_{\theta_0}>0$ must be distinguished  from those in which $(\partial B/\partial\varphi)_{\theta_0}<0$.  This can be done by writing
\begin{eqnarray}
\left(\frac{\partial \varphi}{\partial\theta_0}\right)_{B>}&\equiv& \left(\frac{\partial \varphi}{\partial\theta_0}\right)_{B} \mbox{   when   } \left(\frac{\partial B}{\partial\varphi}\right)_{\theta_0}>0;\hspace{0.2in}\\
\left(\frac{\partial \varphi}{\partial\theta_0}\right)_{B<}&\equiv& \left(\frac{\partial \varphi}{\partial\theta_0}\right)_{B} \mbox{   when   } \left(\frac{\partial B}{\partial\varphi}\right)_{\theta_0}<0.
\end{eqnarray}

The dimensionless function $\mathcal{S}(\theta_0,B_t)$ on each $\psi$ surface can be rewritten as
\begin{eqnarray}
\mathcal{S}&=& N_p B_t\int_{B_{min}}^{B_t}  \frac{\left(1-\frac{B}{2B_t}\right)}{\sqrt{1-\frac{B}{B_t}}}\frac{\left(\frac{\partial \varphi}{\partial\theta_0}\right)_{B>}- \left(\frac{\partial \varphi}{\partial\theta_0}\right)_{B<}}{B^2}dB. \nonumber\\  \label{psi-drift}
\end{eqnarray}

Equation (\ref{psi-drift}) defines a sufficient condition for exact omnigeneity on a magnetic surface $\psi$.  Exact omnigeneity is obtained when the dimensionless quantity $\mathcal{S}(\psi,\theta_0,B_t,)=0$ for every possible value for $B_t$ and $\theta_0$.  This is the case when
\begin{equation}
\left(\frac{\partial \varphi}{\partial\theta_0}\right)_{B>} =\left(\frac{\partial \varphi}{\partial\theta_0}\right)_{B<} \mbox{  at each } (\theta_0,B). \label{Exact-omnig}
\end{equation}

One can use the symmetry in the derivative of $\zeta\equiv N\varphi - M\theta$ instead $\varphi$;
\begin{equation}
\left(\frac{\partial \varphi}{\partial\theta_0}\right)_{B} = \frac{M}{N-\iota M} + \frac{\left(\frac{\partial \zeta}{\partial\theta_0}\right)_{B}}{N-\iota M}.
\end{equation}
In quasi-symmetry, $\big(\partial\zeta/\partial\theta_0\big)_B=0$, and Equation (\ref{Exact-omnig}) is automatically satisfied.


\subsection{The bounce time }

The expression for the bounce time in terms of the action was obtained in Equation (\ref{partial J/ partial u}).  Here, that expression will be expressed in terms of $B_t$ and $B$, and this expression will be evaluated for deeply trapped particles.

\begin{eqnarray}
\tau_b &=& \frac{\partial J}{\partial u} \\
&=& \frac{\mu_0(G+\iota I)}{2\pi} \int \frac{d\varphi}{B\sqrt{\frac{2}{m}(u-q\phi -\mu B)}}\\
&=& \frac{\mu_0(G+\iota I)}{N_p vB_t}\mathcal{D}_\tau, \mbox{  where  } \label{tau_b-D} \\
\mathcal{D}_\tau&\equiv&\frac{B_tN_p}{2\pi}\int_{-\varphi_t}^{\varphi_t} \frac{d\varphi}{B   \sqrt{1-\frac{B}{B_t}}},
\end{eqnarray}
$\varphi_t$ is the turning point of the banana, $N_p$ is the number of periods of the stellarator, and $\mathcal{D}_\tau$ is the dimensionless duration of the time between turning points.

The bounce time depends on the $\varphi$ dependence of the field strength along a field line.  A simple assumption is $B=B_{max}-(B_{max}-B_{min})(1+\cos N\varphi)/2$ with $N$ the number of periods.  Even this assumption gives a complicated integral which is logarithmically singular as $B_t\rightarrow B_{max}$.  Nevertheless, the bounce time near the field minimum can be easily calculated and is representative of the bounce time except for barely trapped particles. Near the minimum has the form $B=B_{min}+(B_{max}-B_{min})N^2\varphi^2/4$ and
\begin{eqnarray}
&& \sqrt{1-\frac{B}{B_t}}= N\varphi_t\sqrt{\frac{\epsilon}{2}}\sqrt{1-\left(\frac{\varphi}{\varphi_t}\right)^2}, \mbox{   where   }\\
&&\epsilon \equiv \frac{B_{max}-B_{min}}{2B_t}. \hspace{0.3in} \mbox{   Since  }\\
&& \int_{-1}^1 \frac{ds}{\sqrt{1-s^2}} = \pi\\
&& \tau_b =\frac{\mu_0(G+\iota I)}{ \sqrt{2\epsilon} N v B_t},   \mbox{   so } \\
&&\mathcal{D}_\tau \approx \frac{1}{\sqrt{2\epsilon}}
\end{eqnarray}
when $\epsilon<<1$.  The major radius $R\equiv \mu_0(G+\iota I)/(2\pi B)$ and the length of a period of a stellarator is $L_p\equiv2\pi R/N$, so the characteristic bounce time for trapped particles is
\begin{equation} \tau_b = \frac{L_p}{\sqrt{2\epsilon} v}.\end{equation}


\subsection{The precession frequency}

The precession frequency, $\omega_{pr}=-2\pi(\partial J/\partial\psi)/q\tau_b$, is determined by the most complicated derivative of the action, $\partial J/\partial\psi$.  This derivative has three terms: one proportional to $d(G+\iota I)/d\psi$, another proportional to $d\phi/d\psi$, and a third proportional to $\partial B/\partial\psi$.  The first term, which is $\left(d\ln(G+\iota I)/d\psi\right)J$, is usually not important in stellarators and will be ignored in the expressions given below for the sake of simplicity.

Differentiating Equation (\ref{full J}) for $J$ with respect to $\psi$ and ignoring $d(G+\iota I)/d\psi$,
\begin{eqnarray}
\frac{\partial J}{\partial \psi} &=& - q\frac{d\phi}{d\psi}\tau_b \nonumber\\&&  -\frac{\mu_0(G+\iota I)}{2\pi}mv \int \frac{ 1-\frac{B}{2B_t} }{B^2\sqrt{1-\frac{B}{B_t} } } \frac{\partial B}{\partial\psi} d\varphi \nonumber\\ \\
&=& - \left( q\frac{d\phi}{d\psi} + mv^2\left<\frac{\partial \ln(B)}{\partial\psi}\right> \right)\tau_b,  \label{J/psi}
\end{eqnarray}
where
\begin{equation}
\left<\frac{\partial \ln(B)}{\partial\psi}\right>\equiv\frac{ \int \frac{ 1-\frac{B}{2B_t} }{B^2\sqrt{1-\frac{B}{B_t} } } \frac{\partial B}{\partial\psi} d\varphi}{ \int \frac{d\varphi}{B   \sqrt{1-\frac{B}{B_t}}} }
\end{equation}

The precession frequency is then
\begin{equation} \omega_{pr} =2\pi \frac{d\phi}{d\psi} +2\pi \frac{mv^2}{q}\left<\frac{\partial \ln(B)}{\partial\psi}\right>. \label{omega_pr}\end{equation}

In a thermal plasma, $mv^2/2$  is independent of the mass of the particle.   When the electron and ion temperatures are equal, the electric and magnetic precession speeds have the same magnitude for both species, but their magnetic precessions are in opposite directions.

The magnetic precession is generally in the opposite direction for deeply trapped, $B_t\approx B_{min}$, and barely trapped particles, $B_t \lesssim B_{max}$.



\section{Construction of omnigenous equilibria \label{costruc-omnig} }

Cary and Shasharina \cite{Cary:1997} noted that the magnetic field strength in a given $\psi$ surface can be parameterized by a dimensionless quantity $\eta$ as $B(\eta)$.  The constant $\eta$ curves must be closed and periodic on the torus.  This ensures a particle with a turning point $B_t(\eta_t)$ can never be carried by its precession in $\theta_0$ to a place on the $\psi$-surface where $B_t>B_{max}$ nor $B_t<B_{min}$.

The periodicity constraint on $\eta$ can be written as
\begin{eqnarray}
\eta&=&\zeta -g(\theta,\eta),  \label{eta-curve},
\end{eqnarray}
where $\zeta \equiv N\varphi - M\theta$.  $N$ and $M$ are integers that have no integer factor other than unity in common.  The Cary-Shasharina function $g(\theta,\eta)$ must be periodic in its arguments.  The form of Equation (\ref{eta-curve}) assumes $N\neq0$.  If $N=0$, then $g$ must be chosen to depend on $\varphi$ and $\eta$ rather than $\theta$ and $\eta$ to have two independent variables.  

The dependence of the Cary-Shasharina function $g(\theta,\zeta)$ on $\theta$ represents the extra freedom of omnigeneity over quasi-symmetry.  Note that when $g$ is a function of $\zeta$ alone, the field is quasi-symmetric.

The function $\mathcal{S}(\theta_0,B_t)$ Equation (\ref{psi-drift}) must be zero for all $\theta_0\equiv\theta-\iota\varphi$ and $B_t$ in a given $\psi$ surface, which implies $(\partial\varphi/\partial\theta_0)_{B>}=(\partial\varphi/\partial\theta_0)_{B<}$ must be satisfied everywhere on the surface.  

This condition will be shown to be equivalent to a condition on $g(\theta,\eta)$,
\begin{eqnarray}
&&\left(\frac{\partial g\big(\theta(\theta_0,\eta),\eta\big)}{\partial\theta_0}\right)_{\eta>} =\left(\frac{\partial g\big(\theta(\theta_0,\eta),\eta\big)}{\partial\theta_0}\right)_{\eta<}  \label{g/theta_0}, \hspace{0.3in}\\\nonumber\\
&&\hspace{1.0in} \mbox{   where   }\nonumber\\ \nonumber\\
&&\theta(\theta_0,\eta) =\frac{\theta_0 + \frac{\iota}{N}\eta + \frac{\iota}{N} g(\theta,\eta) }{1-\frac{\iota}{N}M}. \label{theta-simple}
\end{eqnarray}

To show this, note that $(\partial\varphi/\partial\theta_0)_B=(\partial\varphi/\partial\theta_0)_\eta$,
\begin{eqnarray}
&&\left(\frac{\partial\varphi}{\partial\theta_0}\right)_\eta =\frac{M}{N-\iota M}+ \frac{1}{N-\iota M}\left(\frac{\partial\zeta}{\partial\theta_0}\right)_\eta \label{phi vs zeta der};\\
&&\left(\frac{\partial\zeta}{\partial\theta_0}\right)_\eta = \left(\frac{\partial g\big(\theta(\theta_0,\eta),\eta\big)}{\partial\theta_0}\right)_\eta, \label{partial zeta/partial theta_0}
\end{eqnarray}
which follows from Equation (\ref{eta-curve}).

Equation (\ref{g/theta_0}) has subtleties, which will be illustrated in Sections \ref{sec:B_min omnig} and \ref{sec:B_max omnig}  by the relatively simple case in which $M=0$ and $g$ is sufficiently small that only terms linear in $g$ need to be retained.

The deviation of $(\partial\varphi/\partial\theta_0)_\eta$ from its quasi-symmetric form has the property that its $\theta$ average is zero.  In Section \ref{omig bs} this property will be shown to imply that the departure of an omnigenous equilibrium from quasi-symmetry has no effect on the bootstrap current in the limit of low collisionality.  The first term on the right-hand side of Equation (\ref{phi vs zeta der}) is the quasi-symmetric form for $(\partial\varphi/\partial\theta_0)_\eta$.  Using Equation (\ref{partial zeta/partial theta_0}), the second term in Equation (\ref{phi vs zeta der}) gives the deviation of an omnigenous magnetic field from quasi-symmetry.  This deviation has a $\theta$-average that is zero since
\begin{eqnarray}
0&=&\frac{\oint \left(\frac{\partial g}{\partial\theta}\right)_\eta d\theta}{N-\iota M}\\
&=& \oint \Big( \left(\frac{\partial\varphi}{\partial\theta_0}\right)_\eta-\frac{M}{N-\iota M}\Big)\left(\frac{\partial \theta_0}{\partial\theta}\right)_\eta d\theta. \label{omnig-integra}
\end{eqnarray}


\subsection{Omnigeneity near $B_{min}$ \label{sec:B_min omnig}}

By an appropriate choice of the starting points of the angles $(\theta,\varphi)$, a field strength minimum can be taken to be at $\eta=0$.   To simplify the discussion it will be assumed that $M=0$ and that only terms that are linear in $g$ need to be retained.  

$M=0$ stellarators, which are called isodynamic \cite{Nuehrenberg:2010} or equivalently quasi-poloidal \cite{Spong:2005},  are a particularly interesting special case.  The results derived for them are related to the properties of omnigenous stellarators in general.

Calculating to order $\eta^2$,
\begin{eqnarray}
g&=&g_0(\theta) + g_1(\theta)\eta + g_2(\theta)\eta^2 + \cdots \mbox{  so  } \\
\theta &=& \theta_0 + \frac{\iota}{N}\eta +  \frac{\iota}{N}g_0(\theta_0)  \cdots \\
g&=& g_0(\theta_0) + g'_0(\theta_0)  \frac{\iota}{N}\eta + \frac{1}{2}g''_0(\theta_0)\left(\frac{\iota}{N}\eta\right)^2 \nonumber\\ &&
+ g_1(\theta_0) \eta + g'_1(\theta_0)  \frac{\iota}{N}\eta^2 + g_2(\theta_0)\eta^2.
\end{eqnarray}

The required symmetry in $(\partial\zeta/\partial\theta_0)_\eta$ implies $(\partial g/\partial\theta_0)_\eta$ cannot contain any odd powers of $\eta$.  The implication is that 
\begin{eqnarray}
&&g'_1(\theta_0) = - \frac{\iota}{N} g''_0(\theta_0) \hspace{0.3in} \mbox{   and  }\\
&&\left(\frac{\partial\zeta}{\partial\theta_0}\right)_\eta = g'_0(\theta_0) + \left\{g'_2(\theta_0) - \left(\frac{\iota}{N}\right)^2\frac{g'''_0(\theta_0)}{2}\right\}\eta^2  \nonumber\\
\end{eqnarray}
through $\eta^2$ order.  The functions $g_0(\theta)$ and $g_2(\theta)$ are arbitrary, but $g_1(\theta)$ must be chosen to cancel a linear term in $\eta$.

The function $g(\theta,\eta)$ can be expanded to arbitrarily large powers of $\eta$.  When this is done, the coefficients of even powers of $\eta$ are arbitrary, but the coefficients of the odd powers must be chosen to ensure $(\partial g/\partial\theta_0)_\eta$ has no dependence on odd powers of $\eta$.


\subsection{Omnigeneity near $B_{max}$ \label{sec:B_max omnig}}

The discussion in Section \ref{sec:B_min omnig} of omnigeneity near a minimum of a magnetic field can be applied near a field maximum as well in a mirror machine, but not in a torus.  When $\eta=0$ at a minimum, small values of $\eta$, positive and negative, are the two sides of the well.   When the magnetic field has only one well and $B=B_0(1-\epsilon_B \cos\eta)$, successive maxima are at $\eta=-\pi,\pi,$ etc.  Unlike in a mirror machine, $\eta=-\pi$ and $\eta=\pi$ are the same contour in a torus and must have the same deformation $\partial g\Big(\theta(\theta_0,\eta),\eta\Big)/\partial\theta_0$.

To illustrate the subtlety, as before let $M=0$ and calculate only to linear order in $g$.  Equation (\ref{theta-simple}) is then $\theta = \theta_0+(\iota/N)\eta$.  The two sides of the same constant-$B$ contour have $\eta=\pi -\delta\eta$ and $\eta=-\pi+\delta\eta$, or
\begin{eqnarray}
&&\frac{ \partial g\Big(\theta_0 +\frac{\iota}{N}(\pi-\delta\eta), \pi-\delta\eta\Big)}{\partial\theta_0} \nonumber\\&&\hspace{0.4in} = \frac{ \partial  g\Big(\theta_0 -\frac{\iota}{N}(\pi-\delta\eta), -\pi+\delta\eta\Big)}{\partial\theta_0} \hspace{0.3in}
\end{eqnarray}
for all $\theta_0$ and $\delta\eta$.  The periodicity of $g(\theta,\eta)$ in $\theta$ makes this impossible when $\iota$ is not an integer.  

The freedom in picking the coefficients of even powers in $\eta$ in a Taylor expansion of $g(\theta,\eta)$ allows one to make a choice that ensures that $\partial g/\partial\theta_0$ is continuous at $\eta=\pm\pi$, which generally requires $\partial g/\partial\theta_0=0$ and quasi-symmetry there.  But for analyticity, not only must this be true but all derivatives of $g$ with respect to $\delta \eta$ must be continuous, which gives an infinite number of constraints and requires quasi-symmetry of all values of $\eta$.  

The behavior of $g$ at the field maximum is the statement of Cary and Shasharina \cite{Cary:1997}   that not only must the magnetic field have the quasi-symmetric form $B_{max}(N\varphi-M\theta)$ at its maximum, but also any deviation from the quasi-symmetry from for values of $\eta$ away from the maximum requires the magnetic field strength be non-analytic.

The lack of analyticity is not surprising given that magnetic field must have the exact quasi-symmetric form at the surface maximum for all values of $\theta_0$, but the extent of the implied deviations from omneigeneity are unclear as are the effects on transport at low collisionality.  As discussed in Section \ref{action failure} small but short wavelength perturbations often prevent action conservation near maxima of the field strength but need not have an unacceptable effect on confinement. 

Assuming action is conserved, Section \ref{departures} will examine the effects of known departures from omnigeneity on transport.



\section{Effects of trapped particle drifts \label{Particle drift}} 

\subsection{General relations \label{General relations}}

Equation (\ref{Exact-omnig}) for exact omnigeneity can be understood using the explicit radial drift to obtain the deviation $\delta\psi_b$ of particles from the $\psi$ surfaces between turning points.  The deviation $\delta\psi_b$ is called the banana width because of the shape of the trajectories.

The drift velocity is given by Equation (\ref{v_d}), and the drift velocity across a magnetic surface that encloses a toroidal flux $\psi$ is
\begin{eqnarray}
&& \vec{v}_d\cdot\vec{\nabla}\psi = \frac{mv^2}{qB^2} \vec{B}\times\Big(\frac{v_{||}^2}{v^2}\vec{\kappa}+\frac{v_\bot^2}{2v^2}\vec{\nabla}\ln B\Big)\cdot\vec{\nabla}\psi, \hspace{0.3in}\\
&& (\vec{B}\times\vec{\kappa})\cdot\vec{\nabla}\psi = (\vec{B}\times\vec{\nabla}\ln B)\cdot\vec{\nabla}\psi,  \\
&& \vec{v}_d\cdot\vec{\nabla}\psi = \frac{mv^2}{qB^2} \left(1-\frac{\lambda B}{2}\right)(\vec{B}\times\vec{\nabla}\ln B)\cdot\vec{\nabla}\psi, \label{radial drift}
\end{eqnarray}
where $\vec{\kappa}\equiv \hat{b}\cdot\vec{\nabla}\hat{b}$ is the curvature of a magnetic field line.  The electric potential has been assumed to be constant on the magnetic surfaces, $\phi(\psi)$.

The first order in $\rho$ change in the $\psi$ position of a particle is given by $d\psi =(\vec{v}_d\cdot\vec{\nabla}\psi)dt$, where $dt=d\ell/v_{||}$.  Since $\vec{B}\cdot\vec{\nabla}B=B\partial B/\partial \ell$ along a line, $d\ell = (B/\vec{B}\cdot\vec{\nabla}B)dB$ along the line. Consequently, the change in $\psi$ as the field strength $B(\psi,\theta_0,\varphi)$ changes on a particular magnetic field line $(\psi,\theta_0)$ is
\begin{eqnarray}
\left(\frac{\partial\psi}{\partial B}\right)_{\theta_0} &=& \frac{mv}{qB^2} \frac{\left(1-\frac{\lambda B}{2}\right)}{\sqrt{1-\lambda B}}
\frac{(\vec{B}\times\vec{\nabla}\ln B)\cdot\vec{\nabla}\psi}{\vec{B}\cdot\vec{\nabla}B} \hspace{0.2in}\\
&=& \frac{mv}{qB^2} \frac{\left(1-\frac{\lambda B}{2}\right)}{\sqrt{1-\lambda B}} Y \label{psi-variation}\\
&& \mbox{ where }\nonumber  \\
Y &\equiv& \frac{(\vec{B}\times\vec{\nabla}B)\cdot\vec{\nabla}\psi }{ \vec{B}\cdot\vec{\nabla}B } \label{H-M form}\\
&=&\mu_0 I -\frac{\mu_0(G+\iota I) \left(\frac{\partial B}{\partial \theta_0}\right)_{\varphi}}{\left(\frac{\partial B}{\partial\varphi}\right)_{\theta_0}},\label{Y-Boozer}\\
&=& \mu_0 I + \mu_0(G+\iota I) \left(\frac{\partial\varphi}{\partial\theta_0}\right)_B\\
&=& Y_{qs} + Y_o.
\end{eqnarray}
Equation (\ref{psi-variation}) implies the deviation of a particle from a $\psi$ surface scales as $mv/qB$, which is its gyroradius.

Using Equation (\ref{phi vs zeta der}) the part of $Y$ that is equivalent to that in a quasi-symmetric system $Y_{qs}$ and the additional part $Y_o$ that arises from the system being omnigenous but not quasi-symmetric are
\begin{eqnarray}
Y_{qs}&\equiv&  \mu_0 I +\frac{ \mu_0(G+\iota I)M}{N-\iota M} \hspace{0.1in}\mbox{and  } \label{Y_qs}\\
Y_o &\equiv&   \frac{\mu_0(G+\iota I)}{N-\iota M} \left(\frac{\partial\zeta}{\partial\theta_0}\right)_B.
\end{eqnarray} 

  The function $Y$ was defined \cite{Helander-Simakov:2008}  by Helander and Simakov in 2008, and they obtained Equation (\ref{Y_qs}).  The form of $Y$ in Equation (\ref{H-M form}) is identical to the upsilon of Equation (59) in Hall and McNamara's discussion of omnigeneity \cite{Hall-McNamara}.
  

\subsection{Banana orbits and bootstrap current \label{omig bs}}

In 2009, Helander and N\"uhrenberg \cite{Helander-Nuhrenberg2009} showed that for the special case of current-free isodynamic stellarators, which means $M=0$, the bootstrap current in perfect omnigeneity is zero.  Here it is shown that the bootstrap current is given by the quasi-symmetric part of the field strength and not modified by the deviation of an omnigenous stellarator from quasi-symmetry. 

The banana width of a trapped particle as it goes from its turning point at $B_t$ to $B_{min}$ in the direction in which $(\partial B/\partial\varphi)_{\theta_0}$ is positive
\begin{eqnarray}
\delta\psi_{b>} &\equiv&\int^{B_{min}}_{B_{t}} \left(\frac{\partial\psi}{\partial B}\right)_{\theta_0} dB\\ &=& \int^{B_{min}}_{B_{t}}  \frac{mv}{qB^2} \frac{\left(1-\frac{\lambda B}{2}\right)}{\sqrt{1-\lambda B}} (Y_{qs}+Y_o)dB \hspace{0.2in} \\
&=&\delta\psi_{b,qs} + \delta\psi_{b,0}.
\end{eqnarray}
The banana width is the sum of a part that is equivalent to that of a quasi-symmetric system $\delta\psi_{b,qs}$ and a part $\delta\psi_{b,o}$ that is due to the system being omnigenous but not quasi-symmetric.  The two parts are
\begin{eqnarray}
&& \delta\psi_{b,qs} =  -\mu_0\left(I+\frac{MG+NI}{N-\iota M}\right)\frac{mv\sqrt{1-\frac{B_{min}}{B_t}}}{qB_{min}}; \nonumber\\   \\
&& \delta\psi_{b,o}=-\frac{\mu_0(G+\iota I)mv}{qN_pB_t} \mathcal{S}_>,  \mbox{   where   } \\
&& \mathcal{S}_> \equiv N_pB_t\int_{B_{min}}^{B_{t}}  \frac{\left(1-\frac{ B}{2B_t}\right)}{B^2\sqrt{1-\frac{B}{B_t}}} \left(\frac{\partial \zeta}{\partial\theta_0}\right)_{B>}dB. \nonumber\\
\end{eqnarray}

Equation (\ref{omnig-integra}) implies that the average of contribution of $\rho_{b,o}$ to the banana orbit width vanishes when integrated over the full range of $\theta_0$.  Since the bootstrap current is due to barely trapped particles being systematically outside of their home flux surface on one leg of the trajectory and systematically inside on the other leg, the omnigenous contribution to the banana orbit does not contribute to the bootstrap current.

The bootstrap current in an omnigenous magnetic field equals the value that would have had had the field been quasi-symmetric, $B(\psi, N\varphi-M\theta)$.


\subsection{Systematic drift of trapped particles \label{systematic drift} }

The banana width of a trapped particle as it goes from its turning point at $B_t$ to $B_{min}$ in the direction in which $(\partial B/\partial\varphi)_{\theta_0}$ is negative, $\delta\psi_{b<}$ is the same except the $\mathcal{S}_>$ is replaced by
\begin{equation} 
\mathcal{S}_<\equiv N_pB_t\int_{B_{min}}^{B_{t}}  \frac{\left(1-\frac{B}{2B_t}\right)}{B^2\sqrt{1- \frac{B}{B_t}}} \left(\frac{\partial \zeta}{\partial\theta_0}\right)_{B<}dB. 
\end{equation}

The turning points of a trapped particle are on the same $\psi$ surface only when $\mathcal{S}_<=\mathcal{S}_>$.   The difference $\delta\psi_{b>}-\delta\psi_{b<}$ gives the radial drift per bounce, which is also given by Equation (\ref{psi-drift}) for $\mathcal{S}=\mathcal{S}_>-\mathcal{S}_<$ since the quasi-symmetric part of the banana motion always obeys the required symmetry.


\subsection{Effects of departures from omnigeneity \label{departures}}

The importance of departures from exact omnigeneity on the transport of plasma depends on (1) at what value of $B$ the departure occurs and (2) the collisionality.   Here it will be assumed that the action is conserved exactly.

\subsubsection{Collisionless trajectories}

The effect of a departure from omnigeneity on collisionless trajectories is described by $\partial \psi/\partial \theta_0=\psi'(\psi,\theta_0,u)$, where $\psi'=-(\partial J/\partial\theta_0)/(\partial J/\partial\psi)$,  Equation (\ref{d psi/d theta}). 

Equation (\ref{J-theta_0-phi}) for $\partial J/\partial\theta_0$ can be rewritten as
\begin{eqnarray}
&&\psi' = - mv^2\left<\frac{\partial \ln(B)}{\partial\theta_0}\right> \tau_b  \mbox{   with  } \label{J-theta_0}; \\
&&\left<\frac{\partial \ln(B)}{\partial\theta_0}\right>\equiv\frac{ \int \frac{ 1-\frac{B}{2B_t} }{B^2\sqrt{1-\frac{B}{B_t} } } \frac{\partial B}{\partial\theta_0} d\varphi}{ \int \frac{d\varphi}{B   \sqrt{1-\frac{B}{B_t}}} }.
\end{eqnarray}
Using Equation (\ref{J/psi}),
\begin{eqnarray}
\psi'(\psi,\theta_0,u)&=&-\frac{\left<\frac{\partial \ln(B)}{\partial\theta_0}\right>}{\frac{q}{mv^2}\frac{d\phi}{d\psi}+\left<\frac{\partial \ln(B)}{\partial\psi}\right>}. \label{radial excursion}
\end{eqnarray}

As discussed in Section \ref{sec:omnngeneity}, the deviation $\Delta\psi$ of a particle from its home flux surface is given by half of maximum minus the minimum value of the integral $\int \psi' d\theta_0$ provided the precession, which is represented by the denominator of Equation (\ref{radial excursion}), does not vanish.  

The precession frequency is given by Equation (\ref{omega_pr}).  For particles with energies comparable to or below the thermal energy $mv_t^2/2 = 3T/2$, the derivative of electric potential, $d\phi/d\psi$, tends to dominate the precession while at higher energies the derivative $\partial \ln(B)/\partial\psi$ dominates.  

The potential $\phi$ is generally set by the need to balance the pressure gradient of the poorer confined species, which makes $|q\phi|\approx T$.  

The variation in the magnetic field strength is determined by the curvature $\kappa$ of the magnetic field lines or the displacement of the magnetic field by the pressure.  The field line curvature has different signs at different places on the magnetic surface, which means the precession produced by the magnetic field generally has different signs for deeply and barely trapped particles and passes through zero for some turning-point value.  For thermal particles, the precession due to $d\ln(B)/d\psi$ is smaller by a factor $a\kappa$ relative to the drift due to $d\phi/d\psi$.  For high energy particles, $mv^2>>T$ the precession due to $d\phi/d\psi$ is negligible.

When the denominator of Equation (\ref{radial excursion}) vanishes, the excursions are especially large with $\delta\psi^2$ given by the integral of $2(\partial J/\partial\theta_0)/(\partial^2J/\partial\psi^2)$ with respect to $\theta_0$.     For alpha particle confinement it is important to ensure omnigeneity, $\partial J/\partial\theta_0$ holds for that turning-point value.

The effect of departures from omnigeneity on the confinement of DT-produced $\alpha$ particles differs significantly from that of thermal ions; high-energy $\alpha$ particles slow down preserving their pitch.  When their collisionless excursion from their birth $\psi$-surface is sufficiently small to avoid striking the wall, their confinement is adequate for fusion.


\subsubsection{Effects of collisions}

 The importance of collisionality depends on the mean free path  $\Lambda_{mfp}$, the characteristic distance a particle goes before undergoing a $90^o$ scatter within its $v_{||}/v$ cosine function.  $\Lambda_{mfp}$ is similar for electrons and ions, approximately 10~km in a power plant, and enters results in two ways: (1) as the ratio of $\Lambda_{mfp}/L_p$, where $L_p$ is the length of a period and (2) as the ratio of $\rho\Lambda_{mfp}/L_p^2$.  The gyroradius $\rho$ differs by the square root of the mass ratio $\rho_i/\rho_e =\sqrt{m_i/m_e}$, which is almost two orders of magnitude. 

The collsionality of omnigenious plasmas depends on the ratio of the precession frequency to the effective collision frequency $\nu_{eff}=\nu_c/\sqrt{\epsilon_B}$, where $\nu_c$ is the collision frequency and $\epsilon_B = (B_{max}-B_{min})/(B_{max}+B_{min})$.  The Fokker-Planck collision operator is diffusive so the time it takes to scatter through all trapped particle pitch angles is $\sqrt{\epsilon_B}/\nu_c$.

Simple step-size arguments imply that the diffusion coefficient of a particle in $\psi$ is given by 
\begin{eqnarray}
D&\approx& \left(\frac{d\psi}{dt}\right)^2 \frac{\nu_{eff}}{\nu_{eff}^2 +\omega_{pr}^2}  \mbox{   where   } \label{D-O} \\
\frac{d\psi}{dt} &=& \frac{2\pi}{q} mv^2 \frac{\mathcal{S}}{\mathcal{D}_\tau}, \mbox{   using  }\\
\frac{d\psi}{dt}&=&-\frac{2\pi}{q}\left(\frac{\partial u}{\partial \theta_0}\right)_{\psi J}= \frac{2\pi}{q}\frac{\partial J/\partial\theta_0}{\partial J/\partial u},
\end{eqnarray} 
(\ref{J-theta_0-S}), and (\ref{tau_b-D}).  Units can be surprising; $mv^2/q =B\rho v$ has units of magnetic flux divided by time.  The important quantity is $\big<D\big>$, an average over the magnetic surface of a Maxwellian distribution.

The derivation of Equation (\ref{D-O}) is highly simplified, and this equation is only qualitatively correct.  A much more complete and accurate calculation of the transport with small departures from omnigeneity was given \cite{Parra:2022} by Vincent d'Herbemont, Parra, et al in 2022 with references to earlier work.

For ions or electrons, $\rho v_{th}\approx 10^4 (T/B)$ in meters-squared per second when the temperature $T$ is given in units of 10~keV and $B$ in Tesla, so $\omega_{pr}\approx 10^4 T/(B a^2)$ with the minor radius in meters.  The collision frequency for deuterium ions is $\nu_d\approx51 n/T^{3/2}$ with $n$ in units of $10^{20}$m$^{-3}$.  The collision frequency for electrons is $\nu_e\approx4\times10^3 n/T^{3/2}$.  Consequently,
\begin{eqnarray}
\frac{\omega_{pr}}{\nu_d}&\approx 200\frac{T^{5/2}}{Bna^2};\\
\frac{\omega_{pr}}{\nu_e}&\approx 2.5\frac{T^{5/2}}{Bna^2}.
\end{eqnarray}


\section{Discussion \label{Discussion}  }

As discussed in Section \ref{sec:genl-tranp}, the achievement of a self-sustaining DT fusion burn has four subtleties.  

(1) The ion and electron distribution functions must be close to local Maxwellians, $f=f_m e^{\hat{f}}$ with $\hat{f}^2$, when appropriately averaged over velocity space, much smaller than unity.  This constraint gives simple relations among $\hat{f}$, the entropy production per unit volume by collisions, and the products of the transport fluxes and the conjugate gradients that drive the transport.  

(2) The Fokker-Planck equation for the distribution functions $f$ is a high-dimensional advection-diffusion equation, which has the particle trajectories as its characteristics.  An implication is exponential separation of neighboring particle trajectories, which is called chaos, can easily arise and exponentially enhance the rate of entropy production.  The confining magnetic field in a toroidal fusion plasma must be carefully designed to avoid chaotic particle trajectories.  Omnigeneity is the weakest general condition that is adequate.  Even so, the plasma itself can produce fluctuating electric and magnetic fields, often on a ion gyroradius scale, that produce trajectory chaos and can give the gyro-Bohm transport seen in toroidal plasma experiments, which produces entropy at a rate that is independent of particle collisions.  

(3) The balance between gyro-Bohm energy transport and DT power production implies a self-sustained DT fusion burn is easiest to sustain at a plasma temperature of approximately 10~keV and becomes an order of magnitude more difficult when the temperature reaches 35~keV, Figure \ref{fig:gyro-Bohm}.  

(4) The power released by DT fusion that is available for self-sustainment of the fusion burn is due to alpha-heating of the electrons, which through electron-ion collisions heats the ions on an equilibration timescale, $\tau_{eq}(T_e)$.  When the electron temperature is sufficiently large, the equilibration time $\tau_{eq}$ becomes long compared to the ion energy confinement time, $\tau_{Ei}$, and the ions become too cold relative to the electrons to sustain a fusion burn, Figure \ref{fig:T_i}.

As noted, adequate confinement of small gyroradius particles in toroidal magnetic configurations is obtained when the magnetic field satisfies the omnigeneity constraint exactly.  This constraint is simply stated in Boozer coordinates, which are the assumed spatial coordinates in this article. 

Two properties of a magnetic field determine whether it is omnigenous:  (1) All of the constant-$B$ contours on the magnetic surfaces must be unbounded in at least one of the angular coordinates.  (2) The magnetic field strength must have a certain symmetry so that the two turning points of trapped particles lie on the same magnetic surface.  This condition is satisfied when $(\partial \zeta/\partial\theta_0)_{\psi B}$ is the the same function of $B$ when $\vec{B}\cdot\vec{\nabla}B$ is positive and negative; $\zeta\equiv N\varphi-M\theta$.

In exact omnigenity, the field strength at maxima has the form $B_{max}(\psi,M\theta-N\varphi)$.   When this form is analytically extended to apply to all $\theta$ and $\varphi$ and not just at the field maximum, the resulting field strength is quasi-symmetric, $B(\psi,M\theta-N\varphi)$.  An implication is that exact omnigeneity is not consistent with analyticity unless the field is exactly quasi-symmetric.  This was shown by Cary and Shasharina \cite{Cary:1997}, but they also showed analytic fields could be close to omnigeneity but far from quasi-symmetry.

Since exact omnigeneity is in principle not possible except in quasi-symmetry, it is important to study how exact omnigeneity can be broken while minimizing the effect on confinement, which is the subject of Section \ref{departures}.  The fundamental difficulty in obtaining exact omnigneity is at magnetic field maxima, but as in the case of a tokamak with ripple, achieving action conservation close to a field maximum is extremely difficult but apparently not extremely important.

If the omnigeneous constraint were exactly satisfied, departures from Maxwellian distributions would scale as the gyroradius to system size.   It is important to investigate whether omnigenity gives significant additional freedom in the design of stellarators that could be quasi-symmetric.  This is in addition to the use of omnigeneity to design stellarators of the W7-X type, in which $M=0$, that do not have a quasi-symmetric limit.

An important starting point in an investigation of how accurately omnigenity can be achieved is to study a particular magnetic surface that contains a fixed toroidal flux $\psi$ in a curl-free magnetic field \cite{Boozer:2019}.  This surface should contain about half the flux that is expected to be enclosed by the outermost confining magnetic surface since this is where the temperature gradient tends to be largest in most models of fusion plasmas.  The trajectory confinement properties near the magnetic axis of a toroidal plasma are irrelevant since analyticity requires that $\vec{\nabla}p=0$ there for quantities such as pressure, density, or temperature.

The most import point is that stellarator configurations that accurately satisfy omnigeneity can be found.  A recent demonstration was given by Jorge, Plunk, Drevlak, et al \cite{Jorge:2022}.

\section*{Acknowledgements}

Discussions at the June 27 to July 8, 2022 workshop of the Simons Foundation collaboration \emph{Hidden Symmetries and Fusion Energy} motivated this paper.  In particular the author thanks Per Helander, Matt Landreman, Elizabeth Paul, Gabriel Plunk, and Felix Parra for stimulating discussions and pointing out misconceptions that I made in earlier versions.  The author also acknowledges the importance of a comment of Per Helander at the 2022 workshop for stimulating interest in the constraints imposed by equilibration when the plasma heating is electron heating.

This material is based upon work supported by the grant 601958 within the Simons Foundation collaboration ``\emph{Hidden Symmetries and Fusion Energy}" and by the U.S. Department of Energy, Office of Science, Office of Fusion Energy Sciences under Award DE-FG02-95ER54333.

\section*{Author Declarations}

The author has no conflicts to disclose.


\section*{Data availability statement}

Data sharing is not applicable to this article as no new data were created or analyzed in this study.


\appendix


\section{Magnetic fields in toroidal plasmas \label{Boozer-coord} }

The discussion of plasma confinement in plasmas confined on magnetic surfaces is simplified by the use of Boozer coordinates \cite{Boozer coord}.  These coordinates always exist for scalar-pressure plasmas confined on nested toroidal magnetic surfaces and represent the magnetic field in the contra- and co-variant forms
\begin{eqnarray}
2\pi \vec{B} &=& \vec{\nabla}\psi\times\vec{\nabla}\theta+\iota(\psi)\vec{\nabla}\varphi\times\vec{\nabla}\psi; \label{contra} \\
&=& \mu_0I(\psi) \vec{\nabla}\theta +\mu_0 G(\psi) \vec{\nabla}\varphi + \beta_*\vec{\nabla}\psi. \label{cov}
\end{eqnarray}
$I(\psi)$ is the toroidal current within a volume that is enclosed by a magnetic surface that contains a toroidal magnetic flux $\psi$, $G(\psi)$ is the poloidal current that is outside that surface, which is the current that passes through the central hole of that torus, $\iota(\psi)$ is the rotational transform or average field-line twist on that surface, $\beta_*(\psi,\theta,\varphi)$ is proportional to the pressure gradient, $\theta$ is a poloidal, and $\varphi$ is a toroidal angle.  

The relation between the contravariant magnetic field representation of Equation (\ref{Clebsch}) and Equation (\ref{contra}) is simple, $\theta=\theta_0+\iota(\psi)\varphi$.  The field line denoted by $(\psi,\theta_0)$ passes through the point $(\psi,\theta)$ at $\varphi=0$.   The covariant representation in $(\psi,\theta_0,\varphi)$ coordinates is 
\begin{equation}
2\pi \vec{B}= \mu_0 (G+\iota I) \vec{\nabla}\varphi + \mu_0I \vec{\nabla}\theta_0 + B_\psi\vec{\nabla}\psi.  \label{cov-theta_0}
\end{equation}


A number of the derivations use of the theory of general coordinates, which is derived in a two-page appendix in either \cite{Boozer:RMP} or \cite{Boozer:NF-rev2015}.  

The position vector $\vec{x}(\psi,\theta_0,\varphi)$ gives the point in $(x,y,z)$ Cartesian coordinates associated with a point $(\psi,\theta_0,\varphi)$ coordinates.  Essential concepts are the Jacobian $\mathcal{J}$, the orthogonality relations, and the dual relations:
\begin{eqnarray}
&& \mathcal{J} \equiv \left( \frac{\partial\vec{x}}{\partial\psi} \times \frac{\partial\vec{x}}{\partial\theta_0} \right)\cdot \frac{\partial\vec{x}}{\partial\varphi} \\
&& \hspace{0.2in}  = \frac{1}{(\vec{\nabla}\psi\times\vec{\nabla}\theta_0)\cdot \vec{\nabla}\varphi},\\
&&  \frac{\partial\vec{x}}{\partial\psi}\cdot\vec{\nabla}\psi =1 \mbox{   for any coordinate  } \psi,  \label{O1}\\
&&  \frac{\partial\vec{x}}{\partial\psi}\cdot\vec{\nabla}\theta_0 =0 \mbox{   for any coordinates $\psi$ and $\theta_0$,  } \hspace{0.3in} \label{O2}\\
&&  \frac{\partial\vec{x}}{\partial\psi} \times \frac{\partial\vec{x}}{\partial\theta_0} = \mathcal{J}\vec{\nabla}\varphi, \mbox{ and  } \label{D1}\\ 
&&\vec{\nabla}\psi\times\vec{\nabla}\theta_0=\frac{1}{\mathcal{J}}  \frac{\partial\vec{x}}{\partial\varphi}. \label{D2}
\end{eqnarray}
Equations (\ref{O1}) and (\ref{O2}) are called the orthogonality relations, and Equations (\ref{D1}) and (\ref{D2})  are called the dual relations. These equations hold for any cyclic choice of the three coordinates.

Equation (\ref{Clebsch}), $2\pi \vec{B}=\vec{\nabla}\psi\times\vec{\nabla}\theta_0$ implies
\begin{eqnarray}
\vec{B} &=& \frac{\vec{B}\cdot\vec{\nabla}\varphi}{2\pi} \frac{\partial\vec{x}}{\partial\varphi}, \label{B dx/dphi}\\
\mathcal{J} &=& \frac{1}{2\pi \vec{B}\cdot\vec{\nabla}\varphi},  \mbox{   and  }\\
&=& \frac{\mu_0(G+\iota I)}{(2\pi)^2B^2},
\end{eqnarray}
which follows from the product of Equations (\ref{contra}) and (\ref{cov}).  It should be noted that the Jacobian of $(\psi,\theta_0,\varphi)$ coordinates equals that of $(\psi,\theta,\varphi)$ coordinates.


Studies of particle confinement in toroidal plasmas essentially means the confinement of particles near particular $\psi$ surfaces, so only the two in-surface coordinates arise in a subtle way.  The third coordinate can be assumed to be $\psi$, and omitting the explicit statement of this coordinate allows a simplification of the notation.




\cleardoublepage


\begin{thebibliography}{99}

\bibitem{Boozer:RMP} A. H. Boozer, \emph{Physics of magnetically confined plasmas}, Rev. Mod. Phys. \textbf{76}, 1071 (2004): doi10.1103/RevModPhys.76.1071.

\bibitem{Northrop-Teller:1960} T. G. Northrop and E. Teller, \emph{Stability of the Adiabatic Motion of Charged Particles in the Earth's Field},Phys. Rev. \textbf{117}, 215 (1960): doi10.1103/PhysRev.117.215.

\bibitem{Boozer:stell-imp} A. H. Boozer, \emph{Why carbon dioxide makes stellarators so important}, Nucl. Fusion \textbf{60}, 065001 (2020): doi:10.1088/1741-4326/ab87af.

\bibitem{Boozer:NF-rev2015} A. H. Boozer, \emph{Non-axisymmetric magnetic fields and toroidal plasma confinement}, Nucl. Fusion \textbf{55}, 025001 (2015): doi10.1088/0029-5515/55/2/025001.

\bibitem{Aref:1984} H. Aref, \emph{Stirring by chaotic advection}, J. Fluid Mech. \textbf{143}, 1 (1984): 
doi.org/10.1017/S0022112084001233.

\bibitem{Boozer:reconnecton2023} A. H. Boozer, \emph{Magnetic field evolution and reconnection in low resistivity plasmas}, $<$https://arxiv.org/pdf/2212.07487.pdf$>$, (February 2023): doi.org/10.48550/arXiv.2212.07487.

\bibitem{Boozer:quasisymmetry} A. H. Boozer, \emph{Transport and isomorphic equilibria}, Phys. Fluids \textbf{26}, 496 (1983): doi10.1063/1.864166.

\bibitem{Nuhrenberg-Zille:1988} J. N\"uhrenberg and R. Zille, \emph{Quasi-helically symmetric toroidal stellarators}, Phys. Lett. A \textbf{129}, 113 (1988): doi10.1016/0375-9601(88)90080-1

\bibitem{Beidler:2021} C. D. Beidler, H. M. Smith, A. Alonso, et al., \emph{Demonstration of reduced neoclassical energy transport in Wendelstein 7-X}. Nature \textbf{596}, 221 (2021): doi:10.1038/s41586-021-03687-w.

\bibitem{Boozer coord} A. H. Boozer, \emph{Plasma equilibrium with rational magnetic surfaces}, Phys. Fluids \textbf{24}, 1999 (1981): doi10.1063/1.863297.

\bibitem{Boozer:fast-path}  A. H. Boozer, \emph{Stellarators as a fast path to fusion}, Nucl. Fusion \textbf{61}, 096024 (2021): doi:10.1088/1741-4326/ac170f.

\bibitem{Boozer:1980} A. H. Boozer, \emph{Guiding center drift equations}, Phys. Fluids \textbf{23}, 904 (1980): doi:10.1063/1.863080.

\bibitem{Boozer:drift-H} A. H. Boozer, \emph{Time dependent drift Hamiltonian}, Phys. Fluids \textbf{27}, 2441(1984): doi10.1063/1.864525.

\bibitem{Hall-McNamara} L. S. Hall and Brendon McNamara, \emph{Three-dimensional equilibrium of the anisotropic, finite-pressure guiding-center plasma: Theory of the magnetic plasma}, Phys. Fluids \textbf{18}, 552 (1975): doi10.1063/1.861189.




\bibitem{Nuehrenberg:2010} J. N\"uhrenberg, \emph{Development of quasi-isodynamic stellarators},  Plasma Physics and Controlled Fusion \textbf{52}, 124003 (2010): doi10.1088/0741-3335/52/12/124003.


\bibitem{Cary:1997} J. R. Cary and S. G. Shasharina, \emph{Omnigenity and quasihelicity in helical plasma confinement systems}, Phys. Plasmas \textbf{4}, 3323 (1997): doi10.1063/1.872473.

\bibitem{Gibson:1963} G. Gibson, W. C. Jordan and E. J. Lauer, \emph{Particle behavior in a static, asymmetric, magnetic mirror geometry}, Phys. Fluids \textbf{6}, 133 (1963): doi10.1063/1.1724498.

\bibitem{Banana drift} R. Linsker and A. H. Boozer, \emph{Banana drift transport in tokamaks with ripple}, Phys. Fluids, \textbf{25}, 143 (1982): doi10.1063/1.863635.

\bibitem{isodrastic} J. W. Burby, R. S. MacKay, and S. Naik,  \emph{Isodrastic magnetic fields for suppressing transitions in guiding-centre motion} posted on arXiv 24 November 2022, $\big<$https://arxiv.org/pdf/2211.13367.pdf$\big>$: doi10.48550/arXiv.2211.13367.

\bibitem{Catto:2019} P. J. Catto, \emph{Collisional alpha transport in a weakly rippled magnetic field}, J. Plasma Phys. \textbf{85},  905850203 (2019): doi:10.1017/S0022377819000151.

\bibitem{Parra et al:2022} V. d'Herbemont, F. I. Parra, I. Calvo, and J. L. Velasco, J. Plasma Phys. \textbf{88}, 905880507 (2022): doi: 10.1017/S0022377822000897.

\bibitem{Paul:2022} E.J. Paul, A. Bhattacharjee, M. Landreman, D. Alex, J.L. Velasco. and R. Nies, \emph{Energetic particle loss mechanisms in reactor-scale equilibria close to quasisymmetry},
Nucl. Fusion \textbf{62}, 126054 (2022): doi:10.1088/1741-4326/ac9b07.




\bibitem{Bader:2019} A. Bader, M. Drevlak, D. T. Anderson, B. J. Faber, C. C. Hegna, K. M. Likin, J. C. Schmitt, and J. N. Talmadge, \emph{Stellarator equilibria with reactor relevant energetic particle losses}, J. Plasma Phys. \textbf{85}, 905850508 (2019):
doi:10.1017/S0022377819000680.


\bibitem{Spong:2005} D. A. Spong, S. P. Hirshman, J. F. Lyon, L. A. Berry and D. J. Strickler, \emph{Recent advances in quasi-poloidal stellarator physics issues}, Nucl. Fusion \textbf{45}, 918 (2005): doi10.1088/0029-5515/45/8/020.

\bibitem{Helander-Simakov:2008} P. Helander and A. N. Simakov, \emph{Intrinsic Ambipolarity and Rotation in Stellarators}, Phys. Rev. Lett. \textbf{101}, 145003 (2008): doi:10.1103/PhysRevLett.101.145003.





\bibitem{Helander-Nuhrenberg2009} P Helander and J N\"uhrenberg, \emph{Bootstrap current and neoclassical transport in quasi-isodynamic stellarators}, Plasma Phys. Control. Fusion \textbf{51}, 055004  (2009): doi10.1088/0741-3335/51/5/055004.

\bibitem{Parra:2022} V. d'Herbemont, F. I. Parra, I. Calvo, and J. L. Velasco, \emph{Finite orbit width effects in large aspect ratio stellarators}, J. Plasma Phys. \textbf{88}, 905880507 (2022): doi:10.1017/S0022377822000897.

\bibitem{Boozer:2019} A. H. Boozer, \emph{Curl-free magnetic fields for stellarator optimization}, Phys. Plasmas \textbf{26}, 102504 (2019): doi10.1063/1.5116721.

\bibitem{Jorge:2022} R. Jorge, G.G. Plunk, M. Drevlak, M. Landreman, J.-F. Lobsien, K. Camacho Mata, and P. Helander, \emph{A single-field-period quasi-isodynamic stellarator}, J. Plasma Phys. \textbf{88}, 175880504 (2022): doi:0.1017/S0022377822000873.



















 
 
 

 





 \end{thebibliography}
\end{document}